\documentclass[sn-mathphys]{sn-jnl}% Math and Physical Sciences Reference Style
\jyear{2021}%

\theoremstyle{thmstyletwo}%

\theoremstyle{thmstylethree}%

\usepackage{amsmath,amssymb,amsfonts}
\usepackage{graphicx}
\usepackage{textcomp}
\usepackage{xcolor}

\usepackage{amsmath,amssymb,amsfonts}
\usepackage{fancybox}
\usepackage{pdfpages}
\usepackage{comment}
\usepackage{caption}
\usepackage{graphicx}
\usepackage{mathtools, cuted}
\usepackage{lipsum, color}
\usepackage{comment}
\DeclareUnicodeCharacter{00A0}{~}
\usepackage{textcomp}
\usepackage{xcolor}
\usepackage{amsmath,amssymb,amsfonts}
\usepackage{graphicx}
\usepackage{textcomp}

\begin{document}

\title[Step-to-Charge]{Step-to-Charge: mW-scale power transfer to on-body devices for long channel ($> 1m$) with EQS Resonant Human Body Powering}

\author[1]{\fnm{Arunashish} \sur{Datta}}\email{datta30@purdue.edu}

\author[1]{\fnm{Lingke} \sur{Ding}}\email{ding359@purdue.edu}
%\equalcont{These authors contributed equally to this work.}

\author*[1]{\fnm{Shreyas} \sur{Sen}}\email{shreyas@purdue.edu}
%\equalcont{These authors contributed equally to this work.}

\affil*[1]{\orgdiv{School of Electrical and Computer Engineering}, \orgname{Purdue University}, \orgaddress{ \city{West Lafayette}, \postcode{47907}, \state{Indiana}, \country{USA}}}

\abstract{
Current limits of harvested energy in wearables are governed by three fundamental quantities, the physical limits of available energy density in ambient powering, safety limits in intentional powering, and the size of the wearable device. Typical energy harvested, except for solar power in favorable outdoor conditions, ranges from $5 \mu W$ to a maximum of $100 - 200 \mu W$ depending upon the available energy. Further, traditional intentional powering methodologies using ultrasound and radio-frequency either have a severe limitation in range of powering or are inefficient due to high path loss in Non-Line-of-Sight scenarios due to absorption by the body. In this study, we propose a novel approach using the human body, the common medium connecting the wearable devices, as a channel to transfer power. We demonstrate Human Body Powering using ``Step-to-Charge," a first-of-its-kind non-radiative, meter-scale powering methodology using a floor-based source and the human body as the channel to transfer power at lower channel losses to charge and power wearable devices across the whole body. The proposed powering methodology allows more than $2 m W$ peak power to be transferred to a wearable device for $>1m$ channel lengths, which is $> 90X$ greater than the state-of-the-art over previous Human Body Powering attempts. Step-to-Charge enables the powering of a new, extended range of wearable devices across the human body, bringing us closer to enabling battery-less perpetual operation using Human Body Power transfer.
}

%%\pacs[JEL Classification]{D8, H51}

%%\pacs[MSC Classification]{35A01, 65L10, 65L12, 65L20, 65L70}

\maketitle 
\section{Introduction}\label{sec1}
\begin{comment}
 Energy harvesting is pivotal in unlocking the perpetual operation of the ever-growing number of wearable devices, thus improving user experience in applications like activity tracking and continuous health monitoring using healthcare accessories. Current state-of-the-art ambient power harvesting methods are limited by reliable available power as well as the size of the harvester.    
\end{comment}
%________________________________________________________________________
\begin{figure*}[b!]
\centering
\includegraphics[width=\textwidth]{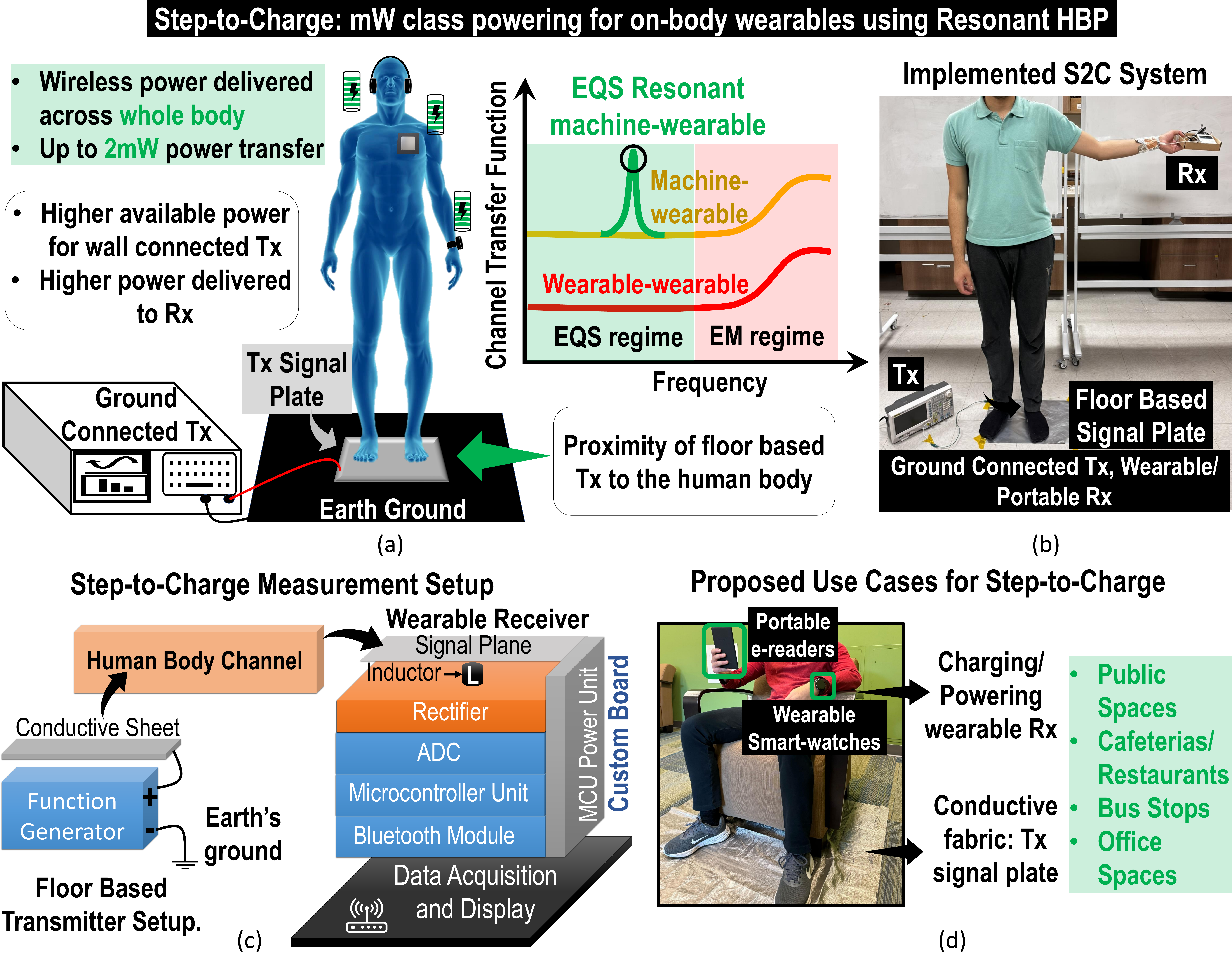}
\caption{\textit{Step-to-Charge: A floor-based, Machine-Wearable wireless charging platform using Electro-Quasistatic Resonant Human Body Powering. (a) A ground-connected charging source transfers power through the body to a wearable receiver. This demonstrates a novel application of EQS Resonant Human Body Powering, which promises low channel loss and higher power delivery. (b) The implemented Step-to-Charge system with a floor—based transmitter was created using Aluminum foil connected to a function generator. (c) A schematic of the Step-to-Charge setup with the transmitter, the human body channel, and the resonant receiver. (d) Step-to-Charge may be used to power portable and wearable devices in public spaces or the gaming and entertainment sector for wireless powering of wearables.}}
\label{fig:intro}
\end{figure*}
%________________________________________________________________________
Wireless powering and energy harvesting methodologies to enable the perpetual operation of remote nodes have been explored for over a century \cite{tesla1904transmission}. In recent times, wireless powering in the context of wearable devices has garnered a lot of interest, with decades of miniaturization of the size of unit computing making wearable technology increasingly accessible. Various energy harvesting methodologies have been investigated for the wireless powering of miniature smart wearable devices. However, the promise of perpetual battery-less operation of wearable devices has not yet been realized. \par
%What are current wireless powering techniques
Harvesting ambient energy in the form of mechanical, thermal, or radiative like piezoelectric, photovoltaic, thermoelectric, or ambient radio frequency (RF) based approaches suffer limitations due to the small size of the harvester as well as sparsely available ambient energy. Availability of ample ambient energy sustained over a long period is restricted by the on-body placement location of the wearable (photovoltaic cell requires light exposure), a size-constrained harvester, and available power density (varying between $5-200 \mu W/cm^3$ for piezoelectric) \cite{chatterjee2023bioelectronic, sanislav2021energy}. Thus, studies have been conducted on intentional powering methodologies using directed radio frequency beams. Wireless charging techniques through radio frequency are restricted by effective charging distance, as in the cases of inductive coupling and low-power density in non-line-of-sight scenarios in microwave radiation \cite{li2021body}. Furthermore, due to the low power transfer efficiency associated with microwave radiation, sufficient received power typically requires a dangerous level of electromagnetic exposure \cite{lu2015wireless}. \par 
%Step-to-charge, description and advantages of using M2W!     
In this paper, we demonstrate Step-to-Charge (Fig. \ref{fig:intro} (a)), where a floor-based earth's ground connected power transmitter is used in a Machine-to-Wearable (M2W) wireless powering architecture to power wearable devices present anywhere on the body. The powering source, being a wall-connected unit, allows us to transmit higher power across a broadband frequency range in the Electro-Quasistatic regime, within the bounds of safety regulations. The large size of the transmitting structure and the proximity to the earth's ground increase the power coupled to the body. Using concepts of Electro-Quasistatic Resonant Human Body Powering (EQS Resonant HBP) \cite{9693115,9431456}, inductive cancellation of parasitic capacitances at operating frequencies of $\leq 30 MHz$ is applied to boost the power received for the wearable devices. The floor-based transmitter also has the advantage of always being close to the human body, and the broadband nature of the powering source allows powering of several receivers simultaneously. We demonstrate the Step-to-Charge system (Fig. \ref{fig:intro} (b)) using off-the-shelf components to receive a peak power of more than $2mW$ through the body for long channel lengths of $> 1m$. We achieve orders of magnitude better power transfer than the state-of-the-art Human Body Powering techniques for wearable devices developed previously for long channel lengths. Step-to-Charge unlocks a large variety of everyday electronics across the whole body, making EQS Resonant Human Body Powering significantly more attractive for daily usage, paving the way for battery-less perpetual operation in wearable wireless Body Area Network devices.

\section{Related Works}

%Using human body as a medium for power transfer and one line for what we are doing
The conductive properties of the human body have been extensively studied as a low path loss alternative to transfer data around the whole body, with numerous applications in communication between the Body Area Network (BAN) devices \cite{zimmerman1996personal,wegmueller2009signal,cho2007human,park2016channel}. The use of the human body as a communication channel at Electro-Quasistatic (EQS) operating frequencies of $\leq 30 MHz$ has been shown to allow communication across the whole body efficiently \cite{sen2020body,maity2018bio,datta2021advanced,maity2020415,maity2021sub,JSSC}. In the EQS frequency regime, the wavelength of the transmitted signals is at least an order of magnitude longer than the dimensions of the human body. In these operating frequencies, the contribution of magnetic fields can be considered negligible \cite{NSR,chatterjee2023biphasic}. In this study, the wearable devices used are capacitive coupling mode devices \cite{maity2018bio} with one of the electrodes connected to the body and the other electrode being a floating ground electrode. It has been shown that the channel loss of a capacitive human body channel is a strong function of the parasitic capacitances between the floating ground of the device and the earth ground, which is the return path capacitance, as well as the capacitance between the floating ground and the human body \cite{nath2019toward, datta2021advanced}.\par

Just like the use of radio frequency (RF) based communication led to the investigation into various methods of RF-based wireless powering, the successful development and demonstration of Human Body Communication prompted the investigation of using conductive properties of body tissues to power on-body devices, which is termed as Human Body Powering. The channel loss for EQS Human Body Powering compares favorably to those of competing wireless power delivery techniques like Radio Frequency or Ultrasound for Non-Line of Sight for long channel length communication around the human body \cite{chatterjee2023bioelectronic,bos2018enabling,ghanbari201917,ghanbari2019sub,thimot2017bioelectronic}, making it an exciting methodology for powering wearable devices.\par

 Previous research efforts on Human Body Powering methodologies have been targeted at improving the effective power received between on-body Human Body Powering (HBP) wearable devices. Having a portable or wearable powering source is convenient for charging wearables on the go. However, a high channel loss for power transfer between wearable devices due to the small return path capacitance ($\leq 1pF$) \cite{datta2021advanced} results in limited power transferred (in the order of a few $\mu W$) when the channel length is of the order of a meter. Thus, even with the state-of-the-art designs, the effective power reliably received for long channels is meager ($\leq 20 \mu W$), \cite{li2021body,li2022body,9693115,9063042,JLi2020,9567745,cho2022intra,lee2022miniaturized,li202034} making this approach viable only for ultra-low-power applications.\par
 In certain specific scenarios where the two wearable devices are in close proximity, the effect of inter-device coupling provides a low path loss channel, resulting in a higher power transfer for small channel lengths \cite{datta2021advanced}. For small channel lengths where the transmitting and receiving devices are close, the floating ground plane of the two devices is capacitively coupled to form a parallel path for the signal loop completion along with the return path capacitances. This results in a reduction in path loss of the body channel for short channel lengths. This reduction in path loss enables a higher power to be transferred to the receiver. Short channel length between the transmitter and receiver is a particular scenario which is typically not guaranteed, and the high power received will not be replicated for longer channel lengths \cite{lee2022miniaturized}. \par 
 Further, previous studies have also used table-top setups to measure power transfer where the transmitter and receiver are both placed on a table, which boosts the return path capacitance through increased coupling to large and close-by environmental objects \cite{cho2022intra}. This reduces channel loss significantly due to a high return path capacitance from the floating ground plate to the table-top, providing a highly optimistic power transfer number. The high power transferred in such a scenario is not sustainable, as such a strong ground coupling is impossible to maintain without large structures close by. Consequently, the transferred power is reduced by orders of magnitude as soon as the setup is tested for a long channel, when worn on the body, or when the transceivers are lifted off the table. Such experiments only target specific scenarios and fail to provide reliable results with increasing channel length for on-body experiments. Most importantly, compared to these studies dealing with the specific optimistic cases, Step-to-Charge can reliably provide higher power transfer across the whole body. \par
 To overcome the limits in capacitive body coupling while adhering to biological and medical safety, this paper takes a novel perspective on a machine-to-wearable wireless power transfer through the body channel. Here, a ground-connected floor-based device is used in conjunction with inductive cancellation of parasitic capacitances to boost the power received. This study provides results for long channel lengths of $>1m$ with true wearable to portable form factor devices to provide realistic results without the optimistic nature of results rising from placing the wearable on a large ground plane or for small channel lengths. \par
    
\subsection{Step-to-Charge: Floor-based Machine-to-Wearable Human Body Powering }
Based on the type of transmitter and receiver used, Human Body Powering (HBP) can be categorized into three categories: \textbf{Machine-Machine (M2M)}, \textbf{Wearable-Wearable (W2W)}, and \textbf{Machine-Wearable (M2W)}. In an M2M scenario, the power source and the receiving device are either connected to the earth's ground or very close to table tops and other large surfaces acting as the earth's ground. Here, the proximity to the earth's ground ensures that the parasitic capacitance between the floating ground of the device and the earth's ground, which forms the circuit's return path, is large, thus reducing the channel loss. However, the use cases of such a system are limited as the receiving device is typically worn on the body or is hand-held, eliminating the proximity to the earth's ground during regular use. For a W2W setup, the transmitting and receiving devices are both wearable devices. In this case, the return path is formed by the combination of the parasitic capacitance from floating ground electrodes of the transmitter and receiver. This makes the channel loss of the setup much higher than M2M cases. \par
An M2W setup is a compromise between the channel loss and portability issues presented by W2W and M2M channels, respectively. As illustrated by the system level figure shown in Fig. \ref{fig:intro} (c), the transmitting power source is an earth's ground connected device, and the receiver is a smaller wearable or portable device on the body. The return path for the setup is formed by the parasitic return path capacitance only from the receiver. The channel loss for an M2W setup is higher than an M2M setup but lower than that observed for a W2W setup, as shown by the concept graph in Fig. \ref{fig:intro} (a). The receiver is a wearable or portable device on the body, and the transmitter is ground connected, enabling a higher power delivered through the body to the receiver than a W2W setup. \par
In this study, we demonstrate Step-to-Charge, a floor-based M2W Human Body Powering methodology that explores the benefits of a ground connected transmitter system to power an extensive range of wearable devices across the body. Step-to-Charge can be used as powering stations in the gaming and entertainment sector or in general day-to-day use in public spaces like cafeterias, bus stops, and office spaces to charge and power on-body devices (Fig. \ref{fig:intro} (d)). \par
\begin{figure*}[b!]
\centering
\includegraphics[width=\textwidth]{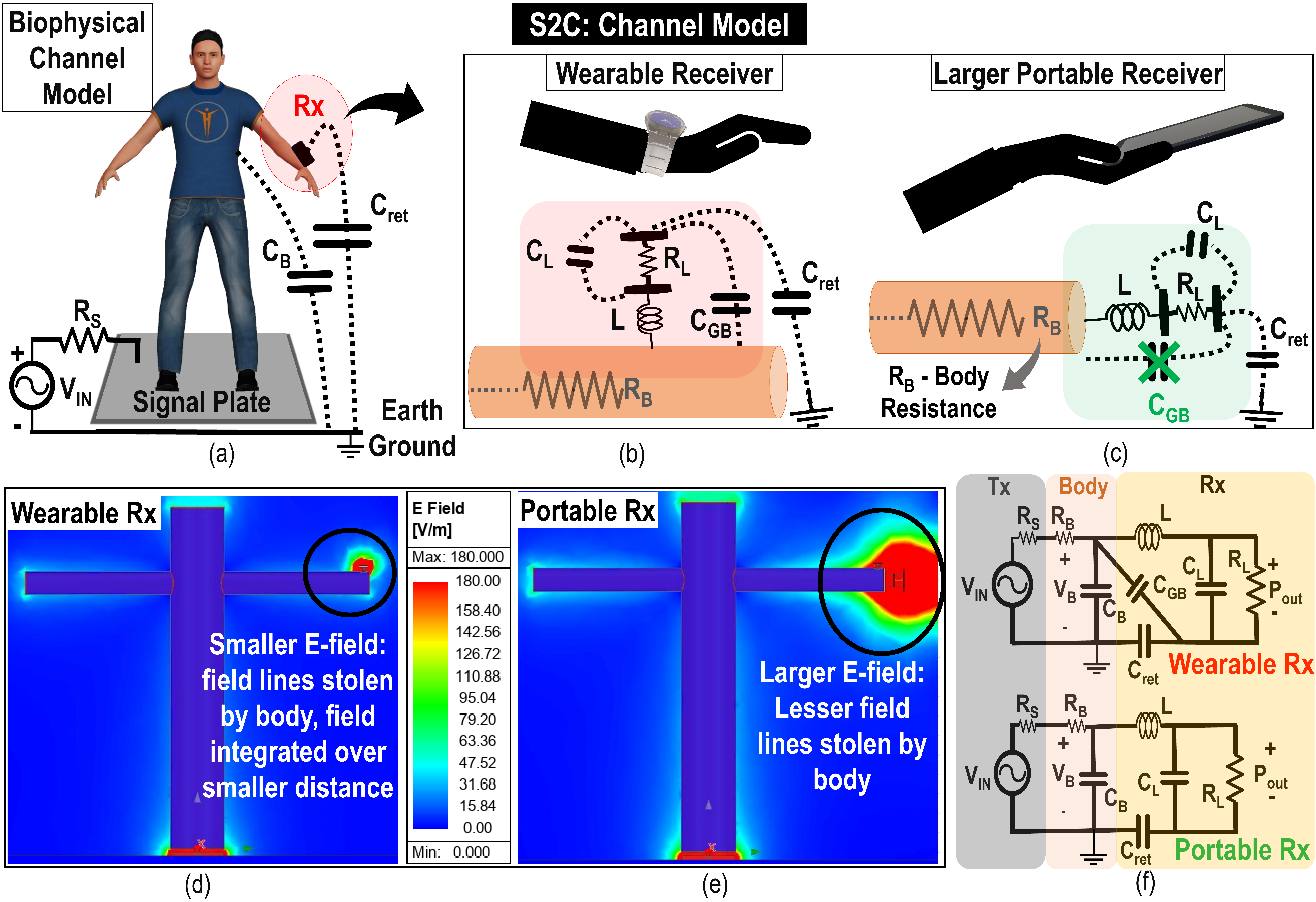}
\caption{\textit{(a) Biophysical model for Electro-Quasistatic Resonant Human Body Powering for Step-to-Charge application. Receivers present on the body are characterized by their size, which are (b) wearable receivers and (c) portable receivers. Depending on the size and location of the floating ground plane of the device, the parasitic capacitances ($C_{ret}$ and $C_{GB}$) change, and their effect on the received power varies. Electric field magnitude is plotted using FEM-based EM analysis tools to show the electric field around (d) wearable small form factor receiver, (e) portable receiver. (f) A circuit model for EQS Resonant Human Body Channel is illustrated for wearable and portable Receivers.}}
\label{fig:HBP_Safety_Study}
\end{figure*}

Step-to-Charge uses the proximity of the floor to the human body at all times to transmit power through the body to any device across the whole body. As a wall-connected power source, the transmitter allows a higher power delivery to wearable receivers than a wearable transmitter. Step-to-Charge further uses inductive cancellation of parasitic capacitances, reducing the channel loss and delivering higher power to the receiver. This Resonant Electro-Quasistatic Human Body Powering uses a series inductor on the receiver to cancel out part of the parasitic return path capacitance ($C_{ret}$ in Fig. \ref{fig:HBP_Safety_Study} (a)). Positioning the transmitting electrode beneath a human's feet leverages gravity to guarantee strong signal coupling to the human body. Further, using the floor as the signal plate to send power through the body is advantageous as it allows unrestricted movement across the room while continuously charging the wearable devices in and around the body.
\begin{comment}
A new range of wearable devices like E-readers and high power physiological sensors, across the whole body, can be powered using Step-to-Charge ushering in the age of smart living spaces and offices. These intelligent rooms and surfaces will allow us to seamlessly charge our wearables when we are in public spaces like cafeterias and bus stops, thus improving the user experience by reducing the worry of our wearable gadgets running out of power. Using the concepts of M2W EQS Resonant Human Body Powering, this paper with experimental results demonstrates that power levels of above $2 mW$ ($\geq 10X$ more than the state-of-the-art) can be delivered to wearable receivers around the whole body.
\end{comment}

\section{Theoretical Analysis}
\label{sec:theory}
A detailed theoretical analysis of Machine-to-Wearable (M2W) Resonant Human Body Powering is performed to understand the potential of Step-to-Charge in powering devices around the body. Biophysical circuit modeling for the human body as a communication channel has been performed previously in literature for EQS human body channels \cite{maity2018bio, datta2021advanced}. In this study, we leverage the concepts of EQS-HBC to develop a simplified circuit model for a Resonant EQS-HBP system. Further, a detailed safety study is performed to analyze the feasibility of using the Step-to-Charge system.  
%________________________________________________________________________
%________________________________________________________________________
\subsection{Electro-Quasistatic Human Body Powering}
\label{sec:theory_start}
Fig. \ref{fig:HBP_Safety_Study} (a) illustrates the biophysical circuit model for an Electro-Quasistatic (EQS) human body channel. In the EQS regime ($\leq 30 MHz$), the wavelength of signals is at least an order of magnitude larger than the dimensions of the human body. This allows us to model the human body as lumped elements per the Quasistatic approximation \cite{NSR,maity2018bio,datta2021advanced}. \par
In the circuit models shown in Fig. \ref{fig:HBP_Safety_Study} (a), the body is modeled by the lumped network formed by the body resistance ($R_{B}$) and the capacitance between the body and the earth's ground ($C_{B}$). The resistive loss due to body tissues is modeled as $R_B$. The forward path for the power transmission is through the body, and the return path is formed by the parasitic capacitance between the floating ground plane of the receiver and the earth's ground, referred to as $C_{ret}$, as shown in Fig. \ref{fig:HBP_Safety_Study} (a). The return path capacitance typically has a magnitude of $\leq 1pF$ for devices of wearable form factors. $C_{ret}$ is a function of the size of the floating ground plane of the device, with a larger floating ground size resulting in a larger $C_{ret}$. Due to the large AC impedance of the return path capacitance, a significant fraction of the voltage drop is across $C_{ret}$, making the voltage across $R_L$ small and thus reducing the received power.\par
 The circuit model is developed for two cases, Fig. \ref{fig:HBP_Safety_Study} (b) shows a wearable receiver, and Fig. \ref{fig:HBP_Safety_Study} (c) shows a portable hand-held device. In the wearable receiver case (Fig. \ref{fig:HBP_Safety_Study} (b)), the parasitic capacitance between the floating ground plate of the device and the body ($C_{GB}$) has a solid contribution to the overall channel loss of the setup along with the return path capacitance ($C_{ret}$). For devices with the same floating ground plane size, being closer to the body increases $C_{GB}$ and worsens the channel loss, resulting in lower received power.\par
For some receivers, typically for portable devices with large floating ground plates, the contribution of $C_{GB}$ is negligible with respect to the return path capacitance $C_{ret}$. This is illustrated by Fig. \ref{fig:HBP_Safety_Study} (c). For receivers that have a high $C_{ret}$ in comparison to $C_{GB}$, the channel loss is a weak function of $C_{GB}$ and in particular limiting cases becomes independent of $C_{GB}$. \par
 This is further motivated by the electric field magnitude plots shown in Fig. \ref{fig:HBP_Safety_Study} (d) and (e). For a wearable receiver (Fig. \ref{fig:HBP_Safety_Study} (d)), the field magnitude is smaller around the receiver owing to a high number of field lines terminating from the floating ground plate of the receiver to the body due to their proximity increasing $C_{GB}$. This also means that fewer electric field lines are terminating to the earth's ground, completing the return path. This reduces the effective return path capacitance ($C_{ret}$) due to higher body shadowing in the case of a wearable small form factor receiver \cite{datta2021advanced}. The field magnitude around the receiver is higher in the case of a larger portable receiver (Fig. \ref{fig:HBP_Safety_Study} (e)). The parasitic return path capacitance is larger due to the larger floating ground plane of the receiver, which is held away from the body (reducing $C_{GB}$) and increasing $C_{ret}$. This motivates the careful design of the floating ground plane to increase $C_{ret}$ while simultaneously reducing $C_{GB}$, which is essential in ensuring high power delivery to the receiver. 

\subsection{EQS-Resonant-Human Body Powering}
\label{sec:resonant_theory}
As discussed in Section \ref{sec:theory_start}, the parasitic return path capacitance ($C_{ret}$) and floating ground plate to body capacitance ($C_{GB}$) make the channel loss high. Thus, for higher usable power (high voltage across $R_L$), reducing the effect of $C_{ret}$ and $C_{GB}$ is vital. This is done by adding a series inductance on the signal path. The use of inductive cancellation to achieve higher power transfer is termed Resonant Human Body Powering.\par
Fig. \ref{fig:HBP_Safety_Study} (f) illustrates the circuit model for the two receiver topologies. Here, $R_S$ is the source resistance of the transmitter supplying $V_{IN}$, $L$ denotes the series inductor added for parasitic capacitance cancelation, and the power received is measured across the load $R_L$ (Eqn. \ref{eqn:Rx_power}).
\begin{equation}
    P_{out} = \frac{V_o^2}{R_L}
    \label{eqn:Rx_power}
\end{equation}

%THIS CAN GO TO THE SUPPLEMENTARY SECTION DIRECTLY!!!
\begin{comment}
The voltage transfer function of the circuit shown in Fig. \ref{fig:HBP_Safety_Study} (f) is given by Eqn. \ref{eqn:resHBC}.
\begin{equation}
    \frac{V_{o}}{V_{B}} = \frac{(j\omega L + Z_L)\frac{1}{j\omega C_{GB}}}{(j\omega L + Z_L).\frac{1}{j\omega C_{GB}} + (j\omega L + Z_L + \frac{1}{j\omega C_{GB}}).\frac{1}{j\omega C_{ret}}} \times \frac{Z_L}{j\omega L + Z_L}
         \label{eqn:resHBC}
\end{equation}    
\end{comment}

Adding the series inductor ensures that a high voltage is observed across the load at the resonant point for a low load resistance ($R_L$), providing us with a higher received power. If we consider the circuit shown in Fig. \ref{fig:HBP_Safety_Study} (f) without the series inductor (L), the output voltage ($V_o$) can be written as shown in Eqn. \ref{eqn:V_o_no_L}. 
\begin{equation}
    V_o = V_B \times \frac{R_L\|Z_L\|Z_{GB}}{Z_{ret} + R_L\|Z_L\|Z_{GB}}
    \label{eqn:V_o_no_L}
\end{equation}
where $Z_{ret}$, $Z_L$, and $Z_{GB}$ are the impedances offered by $C_{ret}$, $C_L$, and $C_{GB}$, respectively.
The impedance of $R_L$ (typically $\leq 1k\Omega$) for optimal power transfer is much smaller than the impedance of $C_L$ and $C_{GB}$ (typically $< 5 pF$) at the operating frequencies ($\leq 10 MHz$). Thus, the parallel combination of the impedances can be simplified, as shown in Eqn. \ref{eqn:impd_simple}.
\begin{equation}
    R_L\|Z_L\|Z_{GB} \approx R_L
    \label{eqn:impd_simple}
\end{equation}
The output voltage across $R_L$ can then be simplified as given by Eqn. \ref{eqn:V_o_simple}.
\begin{equation}
    V_o = V_B \times \frac{R_L}{Z_{ret}+ R_L}
    \label{eqn:V_o_simple}
\end{equation}
For typical values of $C_{ret}$ ($\approx 0.5 pF$) \cite{datta2021advanced}, at a resonant frequency of $1MHz$, which is well within the EQS range, the impedance offered by the return path capacitance ($Z_{ret}$) is $>300 k\Omega$. Thus, the voltage division between $R_L$ and $Z_{ret}$ provides a minimal output voltage unless the load resistance is of the order of $100s$ of $k\Omega$, in which case the received power (Eqn. \ref{eqn:Rx_power}) becomes very small.\par

On adding a series inductor (L), we derive the resonant frequencies and the maximum output voltage for wearable and portable receivers using the biophysical circuit model in Fig. \ref{fig:HBP_Safety_Study} (f). A complete derivation of Eqn. \ref{eqn:resonance}, \ref{eqn:maxVo}, and \ref{eqn:largedevice} is provided in Supplementary Note 1 and Supplementary Figure 1. The closed-form expression for the resonant frequency ($\omega_o$) at which we obtain the maximum output voltage ($V_o$) is given by Eqn. \ref{eqn:resonance}. 
\begin{equation}
    \omega_{o} = \frac{1}{\sqrt{L \times (C_{ret}+C_{GB})}}
         \label{eqn:resonance}
\end{equation}
Further, the output voltage as a function of the body potential ($V_B$) across the body to earth's ground capacitance ($C_B$) at resonant frequency ($\omega_o$) is given by Eqn. \ref{eqn:maxVo}. 
\begin{equation}
    V_o = V_B \times \frac{C_{ret}}{C_{ret}+ C_{GB}}
    \label{eqn:maxVo}
\end{equation}

The output voltage given by Eqn. \ref{eqn:maxVo} is independent of the load resistance ($R_L$), given enough current can be supplied. This ensures higher power is delivered to the receiver at the resonant frequency.\par
Eqn. \ref{eqn:maxVo} further illustrates the benefits of using a portable device where the floating ground plane of the device is held away from the body, as shown by Fig. \ref{fig:HBP_Safety_Study} (e) over a small form factor wearable device (Fig. \ref{fig:HBP_Safety_Study} (d)). A portable device maximizes the potential of a resonant human body channel by increasing $C_{ret}$ with a large ground plate. Moreover, as the ground plane is further away from the body, $C_{GB}$ is small with respect to $C_{ret}$. The resonant frequency and the corresponding voltage at the resonant frequency are given by Eqn. \ref{eqn:largedevice}. 
\begin{equation*}
    \omega_{o} = \frac{1}{\sqrt{L \times C_{ret}}}
\end{equation*}
\begin{equation}
    V_o = V_B
    \label{eqn:largedevice}
\end{equation} 
The output voltage is equal to the body potential at the point the receiver is in contact with the body and independent of the values of other passive elements in the circuit model. The voltage across $R_L$ and, consequently, the received power is maximized at the frequency at which the impedance of the inductor cancels out the impedance of the return path capacitance, giving us a high power for portable receivers. \par

\subsection{Safety Study}
\label{sec:safety}
A safety study is essential to ascertain and analyze the operation range of Step-to-Charge for use by the general public. As illustrated by a previous study investigating the safety aspects of Electro-Quasistatic Human Body Communications (EQS-HBC) \cite{Safety_Study}, the Institute of Electrical and Electronics Engineers (IEEE) \cite{ieee1992ieee} and the International Commission on Non-Ionizing Radiation Protection (ICNIRP) \cite{ICNIRP} have separately recommended the maximum safety limits concerning human exposure to radio frequency (RF) electromagnetic fields. \par
The frequency of operation in Step-to-Charge is $100 kHz - 10 MHz$, which falls well within the EQS regime. ICNIRP guidelines are primarily used to perform the safety study. The ICNIRP guidelines consider people with preexisting medical conditions, metallic implants, and implanted medical devices outside the scope of this study. \par
\begin{figure*}[t!]
\centering
\includegraphics[width=\textwidth]{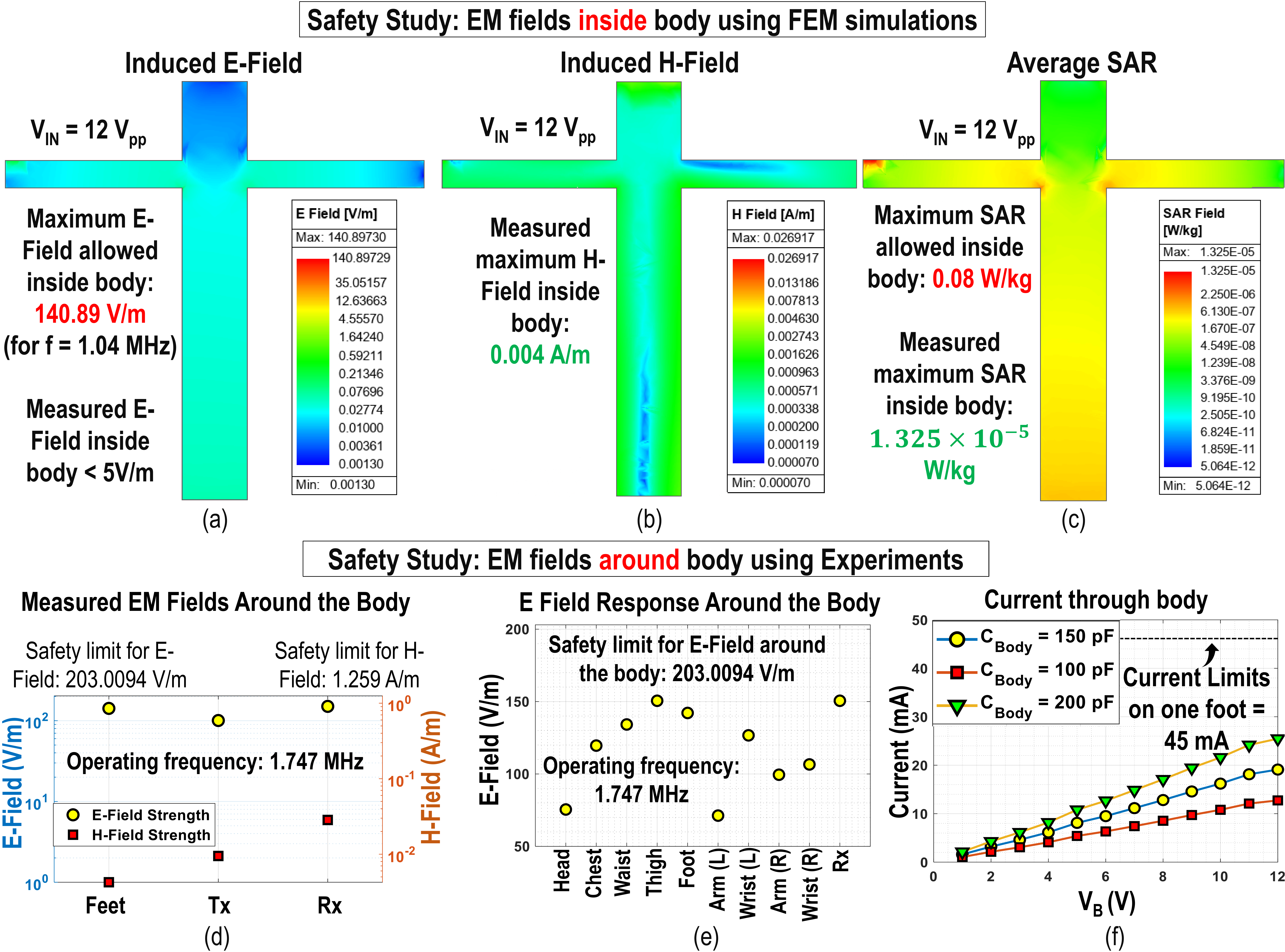}
\caption{\textit{A safety study for Step-to-Charge is demonstrated in this figure for an input voltage of $12V_{pp}$. The ICNIRP guidelines \cite{ICNIRP} are the basis for this safety study. ICNIRP guidelines provide restriction levels on the ``Induced Electric Field" inside the body and the ``Average Specific Absorption Rate" (SAR) which are measured using FEM-based EM simulations on Ansys HFSS. (a) The ``Induced E-Field" inside the body is shown to be orders of magnitude less than the maximum allowed E-Field inside the body. (b) The magnetic field magnitude inside the body is minimal at the operating frequency of $\leq 10 MHz$ as per the quasistatic approximation. (c) The average SAR value is observed to be $> 2X$ orders of magnitude lower than the maximum allowed average SAR as per the ICNIRP guidelines. (d) Experimentally measured E and H Fields around the body, forming the ``reference levels" per the ICNIRP guidelines. (e) A detailed analysis of the E-Field magnitude around the body. (f) Projected contact current for $12V_{pp}$ transmitted signal in case of accidental direct contact with the transmitter signal plate.}}
\label{fig:Safety_Study_Final}
\end{figure*}

The ICNIRP guidelines adopt conservative estimates to provide ``basic restriction" levels, which provide limits on the parameters ``Specific Absorption Rate" (SAR) level and the ''Induced Electric Field" inside the body at the frequency range of interest ($100 kHz$ to $10MHz$). These quantities are difficult to measure experimentally. We use Finite Element Analysis based electromagnetic simulations using Ansys High Frequency Structure Simulator (HFSS) to estimate the values of the quantities (induced electric and magnetic field inside body, Average SAR). Fig. \ref{fig:Safety_Study_Final} (a) and (b) illustrate the induced electric and magnetic field inside the body. We observe that the induced electric field inside the body is $>20X$ less than the maximum allowed by the ICNIRP guidelines. The Average SAR illustrated in Fig. \ref{fig:Safety_Study_Final} (c) also shows that Step-to-Charge is more than 2 orders of magnitude lower than the Average SAR limits imposed by the ICNIRP guidelines. The FEM-based EM simulations show that the induced EM fields and the SAR inside the body are orders of magnitude below the safety limits defined at the input voltage of $12V_{pp}$, ensuring the safe operation of Step-to-Charge.\par
ICNIRP guidelines further define ``Reference levels," which are derived using mathematical approximations and measurements to provide limits on parameters that are easier to observe experimentally. The reference levels, which are based on further conservative approximations on the ``Basic Restrictions," provide constraints on parameters ``Incident Electric Field" and ``Incident Magnetic Field" over the frequency range of interest ($100 kHz$ to $10MHz$). Thus, to further verify the observations in the Finite Element Analysis based EM simulations, we perform experiments to determine the safe operation range of Step-to-Charge. \par
%________________________________________________________________________
%________________________________________________________________________

Fig. \ref{fig:Safety_Study_Final} (d) illustrates the measured incident electric and magnetic field around specific positions on or near the body for an input voltage of $12 V_{pp}$ at an operating frequency of $1.747 MHz$ with the function generator used as the transmitter \cite{Function_generator}. It can be observed that the incident magnetic field is at least an order of magnitude lower than the limits defined by the ICNIRP Reference level guidelines. However, the incident electric field values are closer to the safety limits provided by the reference levels. To analyze the incident electric field around the body further, we perform detailed experiments to observe the variation in the electric field around the body (Fig. \ref{fig:Safety_Study_Final} (e)). This allows us to observe that the reference limits are not crossed at any point around the body at the frequency of operation up to a source voltage of $12 V_{pp}$ on the source meter used. \par
We finally illustrate in Fig. \ref{fig:Safety_Study_Final} (f) that the current through the body is less than the maximum allowed contact current through one foot for an input voltage of $12V_{pp}$. During the experiments, the person standing on the Step-to-Charge platform is not in direct contact with the signal plate. This prevents injuries in case of accidental direct exposure to the signal or the ground plate, ensuring the safety of Step-to-Charge up to an input voltage of $12 V_{pp}$.
   \begin{comment}
    Based on the two standards, HBP operations are restricted by the two most stringent safety limits to prevent tissue heating and electrostimulation. First, the total power transferred from the transmitter to the human body must be less than 0.08W/Kg specific absorption rate (SAR) (6.4W for an average body weight of 80Kg)[citation]. Second, the induced current through each foot must be smaller than 45mA[citation].

    The proposed Step-2-Charge HBP system is strictly below the current limit by which received power of up to 2mW is observed with only up to 20mA of induced body current. As the S2C system requires the user to stand on the charging platform with both feet, this scheme ensures an additional \textcolor{red}{13dB} current safety margin below the safety standard recommendations. \textcolor{red}{Furthermore, due to a more efficient grounded transmitter, the SAR at 20mA is only 2.65uW/Kg for an average adult of 80Kg, significantly below the SAR limit.} Consequently, S2C assures sufficient wearable received power and user safety. 
\end{comment}

\section{Experimental Results}

A floor-based transmitter setup is created using a ground connected function generator connected to an Aluminum foil placed on the floor to form the signal plate. The experiments measure the maximum power that can be received by staying well within the safety standards discussed in Section \ref{sec:safety} using off-the-shelf components.
\par
\begin{figure}[t!]
\centering
\includegraphics[width=\textwidth]{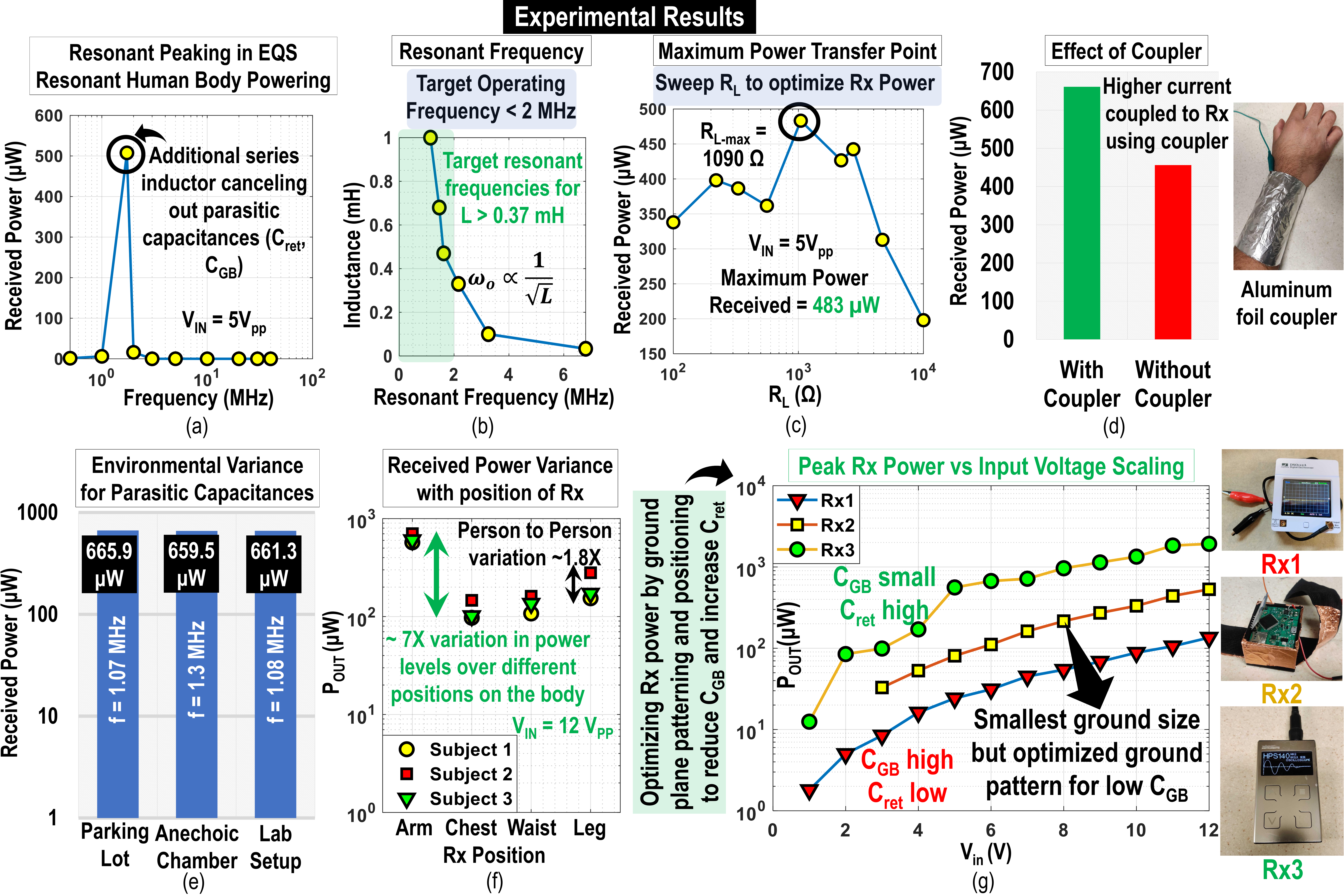}
\caption{\textit{(a) The resonance effect is illustrated with a resonant peak observed at $1.6MHz$ for an inductance of $0.33mH$. (b) The operating frequency can be switched by changing the inductance value to change the frequency of resonance. (c) Power transfer is optimized by sweeping the resistive load till the optimal load resistance is reached. (d) The effect of using a coupler with the receiver to increase the amount of current flowing into the receiver is shown. (e) Variation of received power for a fixed receiver structure in different environments due to changing parasitic capacitances is illustrated (f) The received power is plotted as a function of the position of the receiver on the body for $3$ different subjects to illustrate the variability in received power with changing parasitic capacitances. (g) The received power is illustrated as a function of the input (source) voltage for $3$ types of receivers. The peak RMS received power for the portable receiver (Rx3) was $2.1mW$ for a long channel length ($>1 m$).}}
\label{fig:results}
\end{figure}

\textbf{1. Resonant Peak Detection:}
The first experiment is to observe the resonance theoretically formulated in section \ref{sec:resonant_theory}. The experiment uses an input voltage of $5V_{pp}$ to find the output power for varying frequencies. Fig. \ref{fig:results} (a) illustrates the effect of resonance on the output received power. An inductance of $L = 0.33 mH$ is used to get a resonant peak at a frequency of $1.6 MHz$ for the receiver (Fig. \ref{fig:setup} (b)). 

\textbf{2. Resonant Frequency Selection:}
The value of the series inductance used defines the resonant frequency at which HBP is operated for a low loss channel through the body. The open circuit voltage ($R_L = \infty$) is measured and optimized to find the frequency at which the peak output voltage is observed, thus providing the resonant frequency. Fig. \ref{fig:results} (b) illustrates the effect of changing inductance on the resonant frequency for the receiver. Changing the position and orientation of the receiver changes $C_{ret}$ and $C_{GB}$ and, consequently, the resonant frequency. \par

\textbf{3. Optimizing Received Power:} 
The load resistance is swept across a range of values to ensure maximum power transfer to the receiver, as shown in Fig. \ref{fig:results} (c). The experiment is performed at an input voltage of $5 V_{pp}$. We maximize power transferred for an inductance of $0.33 mH$ with a resonant peak at around $1.6 MHz$. An $R_{L}$ of $1k\Omega$ provides the maximum power, as illustrated by Fig. \ref{fig:results} (c).
\par

%________________________________________________________________________
%________________________________________________________________________

\begin{comment}
\textbf{4. Body Potential:}
The body potential is measured by measuring the signal strength using a ground connected tabletop oscilloscope. The body potential is measured as a function of supplied input voltage by the function generator at a frequency of $1 MHz$. The body potential increases linearly with the input voltage, as illustrated by Fig. \ref{fig:results}. Further, we observe the variance of body potential for a frequency sweep with a fixed input voltage of $5V_{pp}$. A low pass effect is observed with a pole at around $3MHz$. The measured body potential is illustrated in Fig. \ref{fig:results}. \textcolor{red}{Use circuit model to explain occurrence of pole. HFSS Simulation for the same?}\par    
\end{comment}

\textbf{4. Effect of having a coupler for the receiver:} The received power is measured for the same receiver configuration under similar conditions in two scenarios, i) with a coupler, an Aluminum foil coupler wrapped around the arm, and ii) by directly holding the signal terminal without a coupler to collect signal from the body. The coupler design is illustrated in Fig. \ref{fig:results} (d) and further discussed in Section \ref{sec:methods} (Fig. \ref{fig:setup} (e)). The experiment is performed for an input voltage of $12 V_{pp}$, as illustrated in Fig. \ref{fig:results} (d). It can be observed that the presence of a coupler efficiently collects the signal, and more current is coupled to the receiver, thus giving us a higher received power. A large coupler ensures a smaller impedance from the body to the receiver and, hence, a higher current through the receiver.
\par
\textbf{5. Environmental Variation:} The received power is measured for the same receiver configuration for $3$ different environments to observe the effect of varying parasitic capacitance on the power received (Fig. \ref{fig:results} (e)). We perform the measurements for an input voltage of $12 V_{pp}$ in an outdoor open-air parking lot, an anechoic chamber, and the standard lab environment. The change in the resonant frequency in the $3$ scenarios shows a change in the parasitic capacitances due to the change in the environment of the experiments. The received power is very similar in these $3$ cases and is found to be $\approx 660 \mu W$.
\par
\textbf{6. Variation due to Position of Receiver} The power received is a function of $C_{ret}$ and $C_{GB}$. To show the variation in power received due to the changing position of the receiver, we compare the peak power received for the same receiver structure for an input voltage of $12 V_{pp}$ across 4 positions on the body, as illustrated by Fig. \ref{fig:results} (f). We observe a $7X$ variation in the received power with the receiver moving across the body. The experiment was performed on $3$ people, and the person-to-person variation in received power was observed to be $\leq 1.8X$. The person-to-person variance is lesser than the receiver's positional variance. This provides a comparative study using a fixed receiver to analyze the effect of changing parasitic capacitances across different receiver positions on the body and for different people. The variability in parasitic capacitances gives us a variation in the power received for the same receiver structure. However, modifying the structure of the receiver to optimize the parasitic capacitances will provide higher received power than illustrated in Fig. \ref{fig:results} (f).\par
\textbf{7. Power Received vs Input Voltage:}
The power received as a function of the increasing source voltage ($V_{in}$) is illustrated in Fig. \ref{fig:results} (g). Three receivers (Fig. \ref{fig:results} (g)) are optimized to receive the highest possible power. This is done by ensuring that the floating ground plane of the devices is positioned as far from the body as possible to lower $C_{GB}$. Further, dangling wires connecting the inductance and load resistance while ensuring they are away from the body was observed to increase the received power. Three different receivers measure the change in received power, with the input voltage scaled from $1V_{pp}$ to $12 V_{pp}$. 
\begin{itemize}
    \item \textbf{\textit{Rx1: Receiver using a Pocket Oscilloscope}} (Fig. \ref{fig:setup} (b)): The device is thin, with a small distance between the floating ground plane of the receiver and the human body. The surface area of the pocket oscilloscope is $83cm^2$. 
    \item \textbf{\textit{Rx2: Wearable Receiver built using custom board}} (Fig. \ref{fig:setup} (c)): The ground plane is designed to maximize the body to floating ground plane distance and minimize $C_{GB}$. The surface area of the ground plane on the box is approximately $64 cm^2$.  
    \item \textbf{\textit{Rx3: Portable Receiver using a portable oscilloscope}} (Fig. \ref{fig:setup} (d)): The surface area of the hand-held oscilloscope used is $106.5 cm^2$. The portable oscilloscope is held outwards, away from the body, to minimize $C_{GB}$ and maximize $C_{ret}$ with the largest area of the ground plane. 
\end{itemize}
The 3 receivers used in the experiment show the effect of changing the floating ground plane's size and position for the same body posture. Received power of more than $2mW$ is observed for an input voltage of $12V_{pp}$ with \textbf{\textit{Rx3}}, which has a low $C_{GB}$ and the largest ground plane area, resulting in the highest $C_{ret}$. \textbf{\textit{Rx2}} has a peak received power of $531 \mu W$, which is $\approx 4X$ lesser received power. This is expected due to \textbf{\textit{Rx2}} having a smaller ground plane than \textbf{\textit{Rx3}}, resulting in a smaller $C_{ret}$. \textbf{\textit{Rx1}} has a high $C_{GB}$ due to the proximity of the floating ground plane and the body, resulting in a high channel loss and small received power. We observe a peak power of $135 \mu W$, which is $> 10X $ less than that for the portable receiver. This illustrates the necessity for careful design of the floating ground plane, as the custom device with a smaller floating ground had a higher received power than the wearable receiver with a pocket oscilloscope. 

\section{Discussion}\label{sec12}
Step-to-Charge uses a Machine-to-Wearable Resonant Human Body Powering methodology, enabling high power transfer to wearable receivers. Using an earth's ground connected transmitter allows a high body potential (Supplementary Note 2 and Supplementary Figure 2) and thus couples a higher power onto the body. Further, as no transmitter side resonance is employed, the Step-to-Charge transmitter is a broadband low-loss channel (Supplementary Note 3 and Supplementary Figure 3). Using resonance on the transmitter as well as the receiver side, as in the case of conventional wearable-to-wearable resonant human body powering, a very narrow frequency band is available for optimal operation where high power is delivered. Tuning the transmitter resonant frequency to provide optimal power is challenging, as the resonant frequency is a function of parasitic capacitances, which change with time. Since Step-to-Charge relies only on receiver side resonance, delivering higher power can be performed by employing a broadband powering source, ensuring the varying resonant frequency of the receiver doesn't impact the power delivered. Further, a person typically uses multiple wearable devices simultaneously. Using a broadband source, as in the Step-to-Charge case, simultaneously powers numerous receivers. This can not be accomplished with a narrowband resonant wearable transmitting source in a conventional W2W resonant Human Body Powering setup. The advantages of Step-to-Charge have been discussed in detail in Supplemental Note 3, where concept figures (Supplementary Figure 3) describe the benefits of Step-to-Charge using a Machine-to-Wearable Resonant Human Body Powering modality.\par
\begin{figure*}[t!]
\centering
\includegraphics[width=\textwidth]{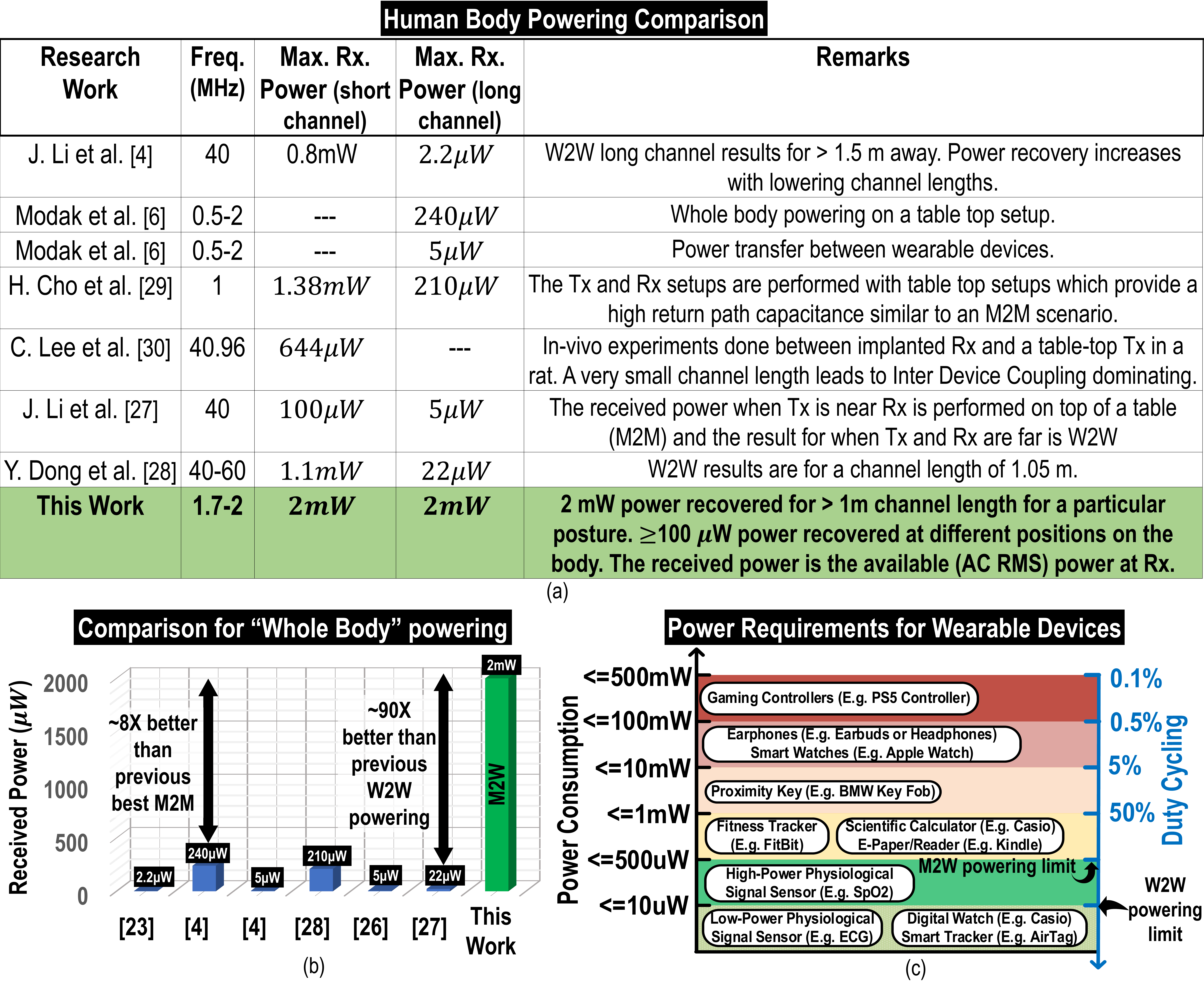}
\caption{\textit{(a) A detailed comparison of Step-to-Charge with previous attempts at Human Body Powering, with results showing the high received power in Step-to-Charge for ``long channel lengths." (b) The performance of Step-to-Charge is compared with other Human Body Powering studies specific to long channel lengths of $\geq 1m$. Step-to-Charge provides $\geq 90X$ better performance than previous state-of-the-art powering wearable devices. (c) The power consumption of typically used wearable and portable devices is illustrated and the ability of Step-to-Charge to charge and power such devices is examined.}}
\label{fig:S2C_Performance}
\end{figure*}

Fig. \ref{fig:S2C_Performance} (a) compares previous works on Resonant Human Body Powering with Step-2-Charge. We compare the results based on frequency of operation and received power for short channel ($\leq 15 cm$) and long channel lengths ($\geq1m$). It can be observed that Step-to-Charge delivers the highest power to wearable and portable devices over long channels. \par

Fig. \ref{fig:S2C_Performance} (b) compares the performance of Step-to-Charge with other Human Body Powering methodologies in powering wearable devices over long channel lengths of $\geq 1m$ proposed previously in the literature. This comparison doesn't consider Machine-to-Machine powering techniques, as the received power is significantly lower for powering wearables. It can be observed that Step-to-Charge offers $>90X$ greater performance than previous attempts at powering wearable devices. Further, Fig. \ref{fig:S2C_Performance} (c) illustrates different types of devices and their typical power consumption levels. We observe that a new class of devices can now be powered (with a conservative estimate of $500 \mu W$ power delivered to a receiver) and charged using Step-to-Charge over reliable wearable-to-wearable powering methodologies. Finally, to demonstrate the developed powering methodology, a hand-held temperature and humidity sensor, ThermoPro TP50 \cite{ThermoPro}, is successfully powered using Step-to-Charge using commercial-off-the-shelf components (Supplementary Video 1 and Supplementary Note 4). This video is available publicly on GitHub \cite{S2C_GitHub}.

%________________________________________________________________________

\section{Conclusion}\label{sec13}
In this study, we demonstrate Step-to-Charge, a Machine-Wearable Electro-Quasistatic Resonant Human Body Powering application capable of delivering mW-scale power to on-body devices across the whole body. We achieve a peak power transfer of $> 2mW$ to an on-body device for long channel lengths of $> 1m$. Step-to-Charge is a novel application that uses a floor-based earth's ground connected transmitter to transfer power through the body to any device connected to the body. We demonstrate $> 90X$ power transferred through the body than the state-of-the-art for Human Body Powering. A theoretical understanding of Step-to-Charge is developed to analyze the EQS Resonant Human Body Powering channel to investigate optimal device design strategies. Using off-the-shelf components, the feasibility of Step-to-Charge was further demonstrated by powering a hand-held temperature and humidity sensor. Step-to-Charge allows us to power a range of wearable devices wirelessly placed across the whole body and ushers in the era of battery-less perpetual operation for wearable Wireless Body Area Network devices.

\section{Methods}\label{sec11}
\label{sec:methods}
\textbf{1. Human Subject Experiments}
The study requiring power transfer measurements to wearable devices with human subjects has been approved by the Purdue Institutional Review Board (IRB Protocol no. IRB-2022-1681). All guidelines and regulations given by the Purdue IRB were followed during the experiments. Informed consent was obtained from all the participants for the experiments.\par
\begin{figure}[t!]
\centering
\includegraphics[width=\textwidth]{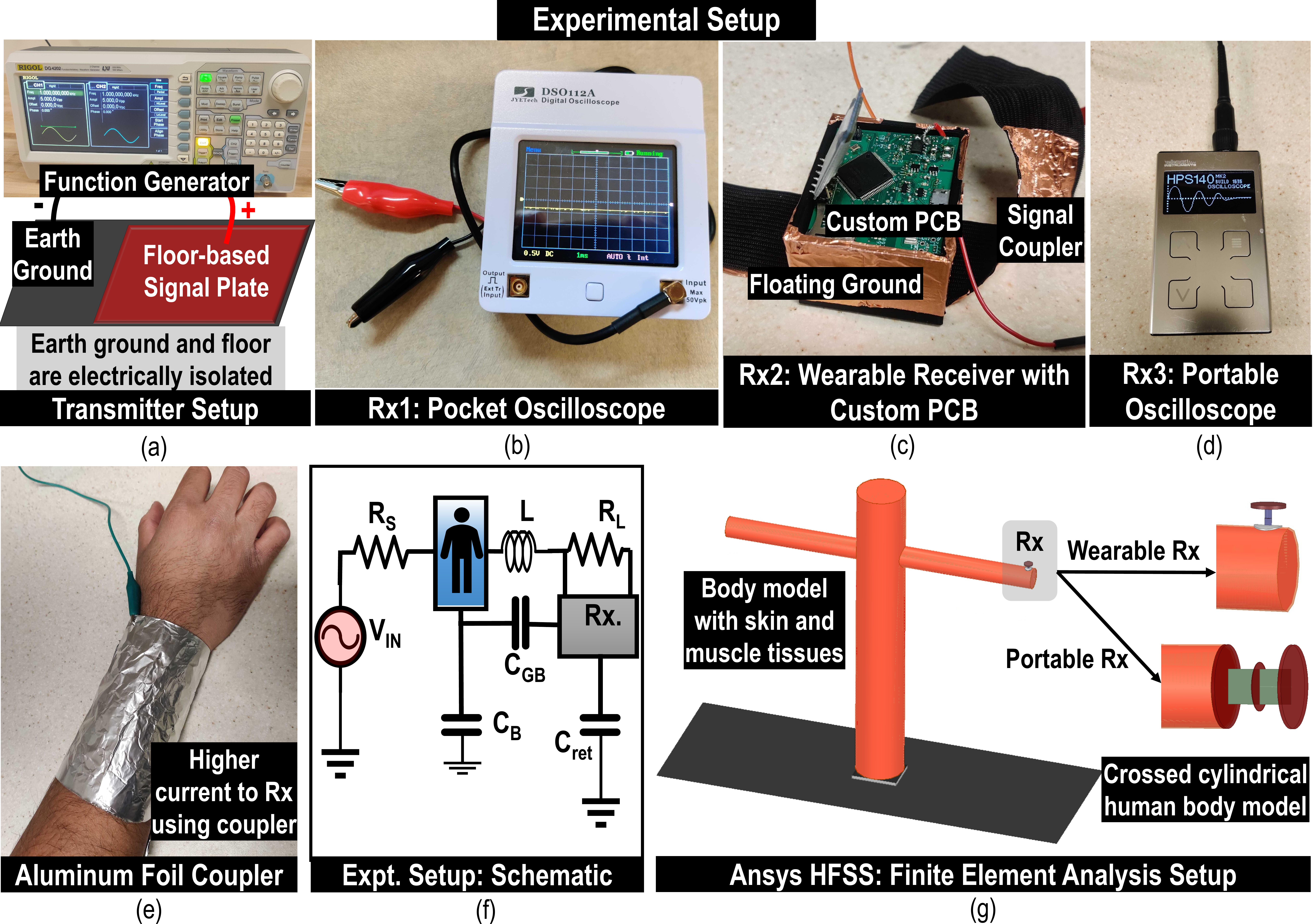}
\caption{\textit{(a) The transmitter setup uses a Rigol, ``DG 4202" Function Generator which is connected to a sheet of Aluminum on the floor which acts as the floor-based signal plate. (b) \textbf{Rx1}: A pocket oscilloscope by JYE Tech, ``DSO Coral 112A Oscilloscope" is used to measure results for small receiver setup. (c) \textbf{Rx2}: A wearable receiver is designed using custom boards housed in a 3D printed box. (d) \textbf{Rx3}: Experiments to measure the power transferred for portable receivers is performed using a hand-held oscilloscope setup ``HPS140MK2" by Velleman. (e) The coupler used in experiments to collect higher current to be delivered to the receiver. (f) The complete experimental setup schematic is illustrated with the important circuit parameters used in the channel modeling. (g) FEM based EM simulations are performed to analyze the safety of Step-to-Charge.}}
\label{fig:setup}
\end{figure}

\textbf{2. Transmitter prototype design:}\\
The transmitter consists of a function generator (Rigol DG4202) \cite{Function_generator} capable of transmitting a sinusoidal wave of peak-to-peak voltage between $0-20$V across a frequency from $0-500 MHz$ (Fig. \ref{fig:setup} (a)), a signal plane on the floor made from aluminum foil and jumper wires connecting the output from the function generator to the aluminum foil. The function generator is plugged into the wall power outlet, connecting the device to the earth's ground. The signal plate is an Aluminum foil sheet placed on the floor of the room, as shown in the schematic in Fig. \ref{fig:setup} (a). The test subject stands on the floor-based transmitter to power the receiver on the body. The test subject is not in direct contact with the signal plane during the experiments, as the experiments are performed with the person wearing a layer of insulation.\par
During these experiments, the floor was observed to be electrically isolated from the earth ground, illustrating that the room floor may not be part of the ``earth ground" to which the function generator is connected. Thus, placing an Aluminum foil sheet on the floor doesn't short the signal plate of the transmitter to the earth's ground. If the floor was a part of the larger earth ground, such a setup may not have worked, as the Aluminum foil would form a low impedance path between the signal and the earth's ground. This would result in a large current flow between the two terminals of the function generator and potentially damage the equipment. In such cases, a layer of insulation must be used (like an insulated carpet), on top of which we may place the Aluminum foil signal plate.\par
%________________________________________________________________________
%________________________________________________________________________
\textbf{3. \textit{Rx1} receiver design using pocket oscilloscope:}\\
The \textbf{\textit{Rx1}} receiver is designed using a small pocket oscilloscope ``DSO Coral 112A Oscilloscope" by JYE Tech \cite{JYETECH}. This has been illustrated in Fig. \ref{fig:setup} (b). The approximate size of the floating ground plane for the system is $83 cm^2$. A coupler made with Aluminum foil wrapped around the hand is connected to a series inductor. The oscilloscope is connected across a load resistance. The voltage measured across the load resistance is used to calculate the received RMS power.\par 
\textbf{4. \textit{Rx2} receiver design using custom board:}\\
Receiver \textbf{\textit{Rx2}} is created using a rectifier connected to a custom PCB, which sends the data using a Bluetooth board (HC05) to a remote data acquisition unit, which displays the received signal. The setup is housed in a 3D-printed box, as shown in Fig. \ref{fig:setup} (c). The ground plane of the setup is wrapped on top of the box to minimize $C_{GB}$ and maximize $C_{ret}$. The size of the ground plane is $64.11 cm^2$. The band tying the box to the wrist has the coupler, which is the signal plate, coupling the signal from the body to the receiver. \par
An inductor is connected in series with the coupler, which is then further connected to a full-bridge rectifier network made from commercial through-hole Schottky diodes (1N5819) with one input left floating. A commercial through-hole capacitor is used after the rectifier network for voltage smoothing before connecting to a load. As the ADC on the custom board has a sampling rate of $480kHz$, measuring the $\leq 2MHz$ signal accurately will not be possible due to aliasing from the severe undersampling. Thus, a rectifier is used to convert the higher frequency signal from the body to DC, which is then measured using the ADC on the custom board. \\
The DC power measured is a function of the power delivered by the body to the receiver and the diode losses in the rectifier. However, characterizing the Human Body Powering channel requires us to calculate the power delivered by the body to the receiver, discounting any extra losses due to non-optimal rectifiers used for measurements. Thus, to find the power delivered to the receiver, we characterize the rectifier to find the input power to the rectifier, which is quoted in this study as $P_{out}$ in Fig. \ref{fig:results} (g). 
\par
\textbf{5. Rectifier characterization:}\\
The full bridge wave rectifier created using Schottky diodes (1N5819) is not optimized to operate with Human Body Powering. The diode losses for the $V_{B}$ observed during the experiments are extremely high, resulting in a significant power loss due to the rectifier. Thus, it becomes essential to characterize the rectifier's power efficiency to find the power delivered by the body to the receiver. \\
The rectifier is characterized by using the same load resistance used during the experiments across it. The function generator is connected to the input of the rectifier with a variable source resistance. The input of the function generator is swept till the output power of the rectifier matches the output power of the wearable receiver used for experiments illustrated in Fig. \ref{fig:results} (g), and the corresponding input power to the rectifier is measured. Varying the source resistance does not change the power efficiency of the rectifier, thus verifying that the input power to the rectifier is the power available to the wearable receiver from the body. \par
\textbf{6. \textit{Rx3} receiver with portable receiver design:}\\
For the portable receiver \textbf{\textit{Rx3}}, the inductor is connected to a battery-operated hand-held oscilloscope, ``Velleman HPS140MK2" \cite{velleman}. The size of the device is $\approx 106.5 cm^2$. The oscilloscope probes are connected across the resistive load to measure the AC voltage received, as shown in Fig. \ref{fig:setup} (d). A coupler is made using aluminum foil wrapped around the test subject's arm. This is connected in series with a commercial through-hole inductor, which is used to cancel out the parasitic capacitances. The RMS power is calculated using the peak-peak voltage observed on the oscilloscope across the resistive load. The complete setup is placed on a box that acts as the portable receiver. \par
\textbf{7. Coupler for receiver:}\\
A coupler is designed using aluminum foil, as illustrated in Fig. \ref{fig:setup} (e). The coupler allows a higher amount of current to be collected from the body to be delivered to the receiver due to a lower impedance presented to the receiver. This ensures that a higher power is available to the receiver through the body channel.\par
\textbf{8. Experimental environment:}\\
The experiments are conducted in a standard lab environment unless otherwise noted. During the experiment, the test subject must stand on the signal plane with no objects or people touching or physically near the test subject. A schematic representation with important circuit parameters is shown in Fig. \ref{fig:setup} (f). During tuning the transmitter resonant frequency, another person turns the frequency dial until the resonant frequency is reached. \par
\textbf{9. Finite Element Analysis based EM Simulations}\\
For safety study analysis and to analyze the behavior of the electric field around the receiver, finite element analysis based electromagnetic simulations are performed using Ansys High Frequency Structure Simulator (HFSS). A simplified human body model is built using skin and muscle tissue properties \cite{Gabriel_1996}. The simplified human body model is made using 2 crossed cylinders, as illustrated in Fig. \ref{fig:setup} (g). The simplified model is used for the simulations to reduce computation time and to minimize simulation effort. Verifying the simplified model used in this study has been performed previously in the literature \cite{Safety_Study} and provides accurate results for wearable devices around the human body. 

\newpage
\section*{Supplementary Notes and Figures}
\hrulefill
\subsection*{Table of Contents}
\textbf{
\begin{itemize}
    \item Supplementary Note 1: Step-to-Charge circuit Modeling and transfer function
    \\
    \item Supplementary Figure 1: Biophysical circuit modeling for Machine-to-Wearable Resonant Human Body Powering
    \\
    \item Supplementary Note 2: Transmitter and receiver design for Step-to-Charge
    \\
    \item Supplementary Figure 2: Floor-based transmitter setup and on-body voltage with an earth's ground connected setup
    \\
    \item Supplementary Note 3: Advantages of using Step-to-Charge
    \\
    \item Supplementary Figure 3: Broadband powering source with a Machine-to-Wearable EQS Resonant HBP setup, allowing multiple receivers to charge simultaneously
    \\
    \item Supplementary Note 4: Demonstration video
    \\
    \item Supplementary Figure 4: A screenshot of the demonstration video illustrating the setup used.
    \\
    \item References
\end{itemize}
}
\newpage
\section*{Supplementary Note 1: Step-to-Charge Circuit Model Transfer Function}\label{sec1}
%________________________________________________________________________
\begin{figure*}[b!]
\centering
\includegraphics[width=0.9\textwidth]{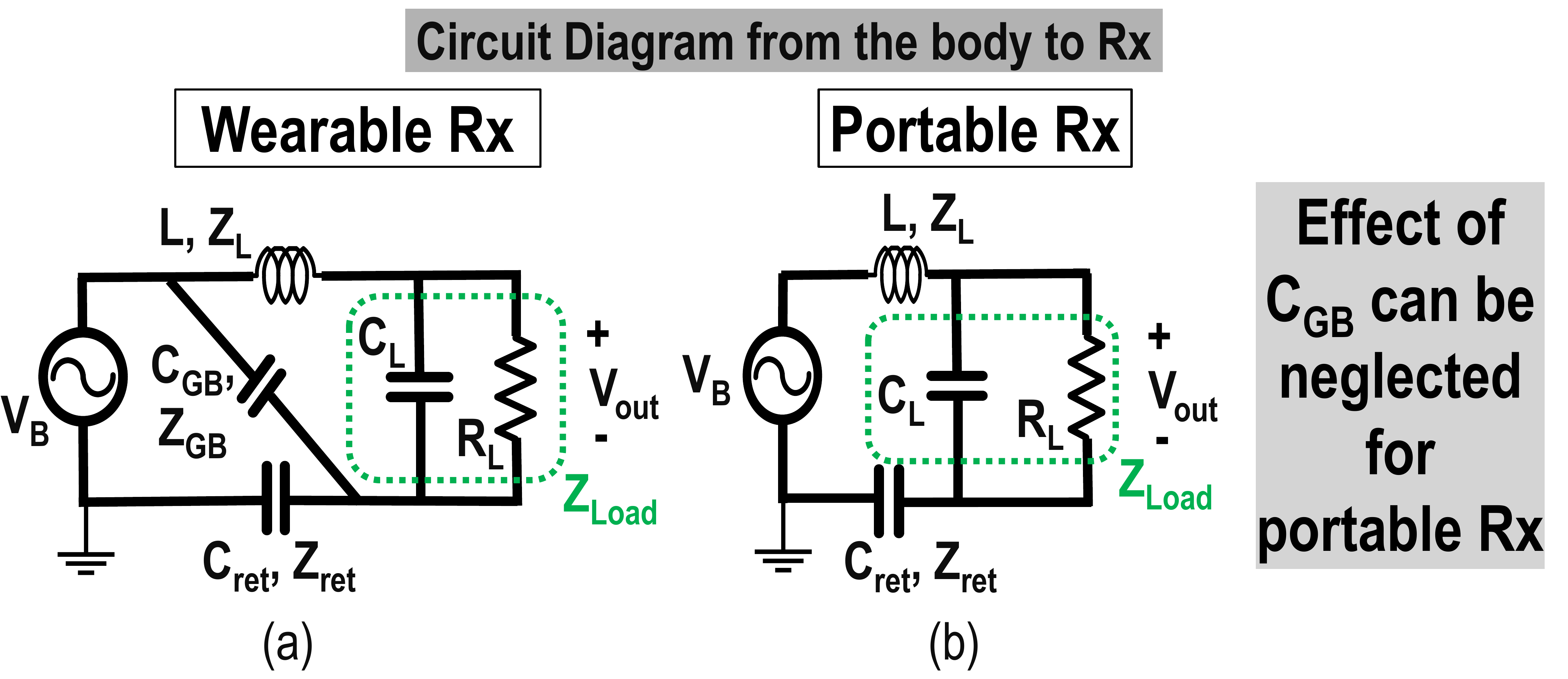}
\caption{\textit{Biophysical circuit modeling of Step-to-Charge channel for (a) wearable and (b) portable receiver.}}
\label{fig:supp_eqn}
\end{figure*}
%________________________________________________________________________
The circuits shown in Fig. \ref{fig:supp_eqn} illustrates a simplified EQS-Resonant Human Body Powering channel from the perspective of the receiver connected to the body \cite{maity2018bio, datta2021advanced, nath2019toward}. $V_B$ denotes the body potential. We solve Fig. \ref{fig:supp_eqn} (a), which is the circuit model for a wearable receiver, and apply simplifications ($C_{GB} \approx 0$) to solve for a portable receiver (Fig. \ref{fig:supp_eqn} (b)).\par

As shown in Fig. \ref{fig:supp_eqn}, we simplify the transfer function by considering the impedance of $L$, $C_{ret}$, and $C_{GB}$ in the circuit model as $Z_L$, $Z_{ret}$, and $Z_{GB}$, respectively. Further, the load is represented as shown in Eqn. \ref{eqn:load}.
\begin{equation*}
    Z_{Load} = R_L\|C_L
\end{equation*}
\begin{equation}
    Z_{Load} = \frac{R_L}{1+j\omega C_L R_L}
    \label{eqn:load}
\end{equation}

The voltage transfer function of the circuit ($V_{o}$/$V_{B}$) shown in Fig. \ref{fig:supp_eqn} (a) is given by Eqn. \ref{eqn:trans_func}.
\begin{equation}
    \frac{V_{o}}{V_{B}} = \frac{(Z_L+Z_{Load})\|Z_{GB}}{(Z_L+Z_{Load})\|Z_{GB} +Z_{ret}} \times \frac{Z_{Load}}{Z_L + Z_{Load}}
         \label{eqn:trans_func}
\end{equation}
Simplifying, 
\begin{equation*}
    \frac{V_{o}}{V_{B}} = \frac{(Z_L+Z_{Load})Z_{GB}}{(Z_{Load}+Z_L).Z_{GB} + (Z_L + Z_{Load} + Z_{GB}).Z_{ret}} \times \frac{Z_{Load}}{Z_L + Z_{Load}}
\end{equation*}  
\begin{equation*}
    \frac{V_{o}}{V_{B}} = \frac{Z_{Load}Z_{GB}}{(Z_{Load}+Z_L).Z_{GB} + (Z_L + Z_{Load} + Z_{GB}).Z_{ret}}
         \label{eqn:resHBC}
\end{equation*}
Simplifying by writing $Z_{GB}$ in terms of the capacitance $C_{GB}$ and frequency ($\omega$),
\begin{equation*}
    \frac{V_{o}}{V_{B}} = \frac{Z_{Load}\frac{1}{j\omega C_{GB}}}{(Z_L + Z_{Load}).\frac{1}{j\omega C_{GB}} + (Z_{Load} + Z_L+ \frac{1}{j\omega C_{GB}}).\frac{1}{j\omega C_{ret}}} 
     \label{eqn:resHBC}
\end{equation*}    
\begin{equation*} 
    \frac{V_{o}}{V_{B}} = \frac{Z_{Load}}{(Z_L + Z_{Load}) + (Z_{Load} + Z_L + \frac{1}{j\omega C_{GB}}).\frac{C_{GB}}{C_{ret}}} 
         \label{eqn:resHBC}
\end{equation*}   
\begin{equation*} 
    \frac{V_{o}}{V_{B}} = \frac{Z_{Load}}{(Z_L + Z_{Load})(1+\frac{C_{GB}}{C_{ret}}) + \frac{1}{j\omega C_{ret}}} 
         \label{eqn:resHBC}
\end{equation*} 
Writing the load impedance ($Z_L$) as shown in Eqn. \ref{eqn:load},
\begin{equation}
    \frac{V_{o}}{V_{B}} =  \frac{R_L}{(j\omega L - \omega^2 L C_L R_L + R_L)(1+\frac{C_{GB}}{C_{ret}}) + \frac{1+j \omega C_L R_L}{j \omega C_{ret}}} 
         \label{eqn:load_simplified}
\end{equation} 
We now write all the impedances ($Z$) in terms of frequency ($\omega$) and the passive components ($L,C_{ret}, C_{GB}$). 
\begin{equation}
    \frac{V_{o}}{V_{B}} =  \frac{R_L}{(j\omega L - \omega^2 L C_L R_L + R_L)(1+\frac{C_{GB}}{C_{ret}}) + \frac{1}{j \omega C_{ret}} + \frac{C_L R_L}{C_{ret}}} 
         \label{eqn:final}
\end{equation} 
Using Eqn. \ref{eqn:final}, we now find the resonant frequency ($\omega_0$). At the resonant frequency, the imaginary part of the denominator becomes $0$. This can be written as \ref{eqn:resHBC}.
\begin{equation}
    j\omega_0 L(1+\frac{C_{GB}}{C_{ret}}) + \frac{1}{j \omega_0 C_{ret}} = 0 
         \label{eqn:resHBC}
\end{equation} 
From Eqn. \ref{eqn:resHBC}, we can find the resonant frequency ($\omega_0$) as shown in Eqn. \ref{eqn:resonance}. 
\begin{equation}
    \omega_{o} = \frac{1}{\sqrt{L \times (C_{ret}+C_{GB})}}
         \label{eqn:resonance}
\end{equation}
The output voltage across the load observed at resonant frequency ($\omega_o$) is found by plugging the value of $\omega_0$ from Eqn. \ref{eqn:resonance} into Eqn. \ref{eqn:final}. This is given by Eqn. \ref{eqn:maxVo}. 
\begin{equation}
    V_o = V_B \times \frac{C_{ret}}{C_{ret}+ C_{GB}}
    \label{eqn:maxVo}
\end{equation}

\newpage
\section{Supplementary Note 2: Transmitter and Receiver Design}

\subsection{Earth's Ground Connected Transmitter}
\begin{figure*}[b!]
\centering
\includegraphics[width=0.8\textwidth]{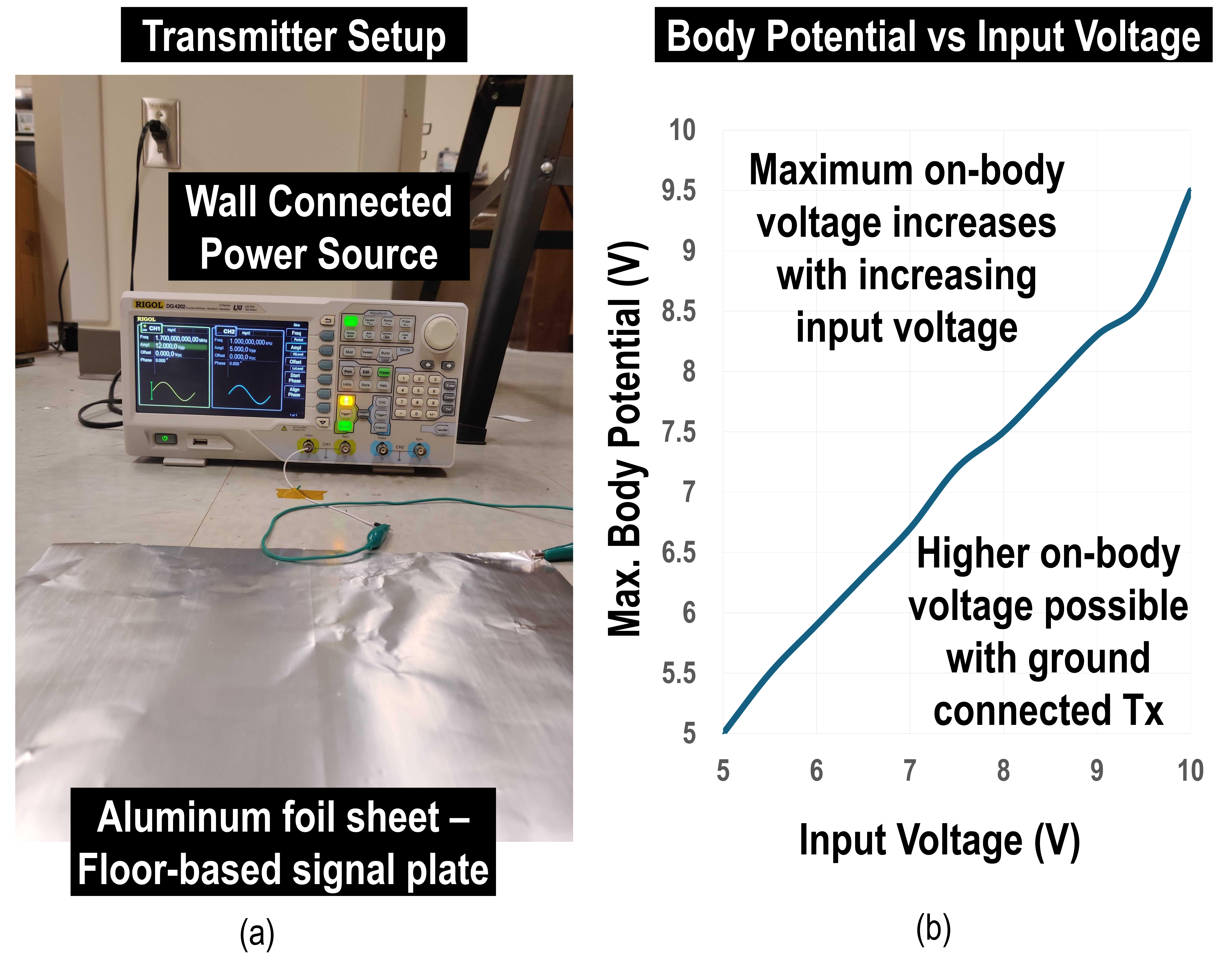}
\caption{\textit{(a) Floor-based transmitter setup used for Step-to-Charge. A wall connected powering source (Function Generator) is used with an aluminum foil sheet used as the signal plate. (b) High on-body voltage ($V_B$) for an earth's ground connected transmitter.}}
\label{fig:Tx_Setup}
\end{figure*}

The transmitter is a power source connected to a floor-based signal plate and the earth's ground. The complete transmitter setup used has been shown in Fig. \ref{fig:Tx_Setup} (a). The transmitter being earth ground connected (wall powered) allows us to deliver higher power through the body within the safety limits defined in Section III of the main manuscript. The maximum on-body voltage illustrated in Fig. \ref{fig:Tx_Setup} (b) shows that high on-body voltages can be obtained for ground connected transmitters.\par
%________________________________________________________________________
%________________________________________________________________________
For wearable transmitters, on-body voltage ($V_{B}$) is given by Eqn. \ref{eqn:on-body_wearable} \cite{maity2018bio,datta2021advanced}.
\begin{equation}
    V_{B} = V_{IN} \times \frac{C_{ret}}{C_B + C_{ret}}
    \label{eqn:on-body_wearable}
\end{equation}
The value of $C_{ret}$ is typically less than $1pF$, and the value of $C_B$ is around $100pF$, as has been verified previously in literature using Finite-Element-Method based EM simulations and experiments \cite{nath2019toward, datta2021advanced}. Thus, we can simplify the on-body voltage equation, as shown in Eqn. \ref{eqn:on-body_wearable_1}. 
\begin{equation}
    V_B \approx V_{IN} \times \frac{C_{ret}}{C_B}
    \label{eqn:on-body_wearable_1}
\end{equation}
Thus, the maximum on-body voltage would be about 2 orders of magnitude less than the input voltage provided by the wearable transmitter (Eqn. \ref{eqn:approx_VB}). 
\begin{equation}
    V_B \approx V_{IN} \times \frac{1}{100}
    \label{eqn:approx_VB}
\end{equation}

This additional loss further results in orders of magnitude lower power available to the receiver using a wearable transmitter. However, in case of an earth's ground connected transmitter, the return path capacitance ($C_{ret}$) is extremely large ($C_{ret} \rightarrow \infty$). Thus, the on-body voltage for earth's ground connected devices can be simplified from Eqn. \ref{eqn:on-body_wearable} to Eqn. \ref{eqn:VB_machine}
\begin{equation}
    V_B \approx V_{IN}
    \label{eqn:VB_machine}
\end{equation}
Fig. \ref{fig:Tx_Setup} (b) shows that the above inference is true with a high on-body voltage observed with an earth's ground connected transmitter. 

\subsection{Receiver Design}\label{sec1}
Optimizing power transfer to the receiver using Step-to-Charge requires careful device design. This can be guided using the theoretical analysis in Section III of the main manuscript. Eqn. \ref{eqn:resonance} and Eqn. \ref{eqn:maxVo} define the key parameters essential in optimizing received power. \par 
\begin{equation}
    \omega_{o} = \frac{1}{\sqrt{L \times (C_{ret}+C_{GB})}}
         \label{eqn:resonance}
\end{equation}
\begin{equation*}
    V_o = V_B \times \frac{C_{ret}}{C_{ret}+ C_{GB}} 
\end{equation*}
\begin{equation}
\frac{V_o}{V_B}= \frac{1}{1+ \frac{C_{GB}}{C_{ret}}} 
    \label{eqn:maxVo}
\end{equation}

From Eqn. \ref{eqn:maxVo}, we observe that reducing the output voltage is a function of the ratio of $C_{GB}$ and $C_{ret}$. The physical design of the receiver is heavily motivated by increasing the return path capacitance ($C_{ret}$) and simultaneously reducing the floating ground plate to body capacitance ($C_{GB}$). This can be achieved by ensuring sufficient distance between the floating ground plane of the receiver and the body, and further ensuring that the size of the floating ground plane is as large as possible to have a high return path capacitance. Strategic design of the floating ground plane to optimize $C_{ret}$ and $C_{GB}$ should be used to maximize the power delivered to the receiver. 
\newpage
\section*{Supplementary Note 3: Advantages of Step-to-Charge}\label{sec1}

%________________________________________________________________________
\begin{figure*}[b]
\centering
\includegraphics[width=\textwidth]{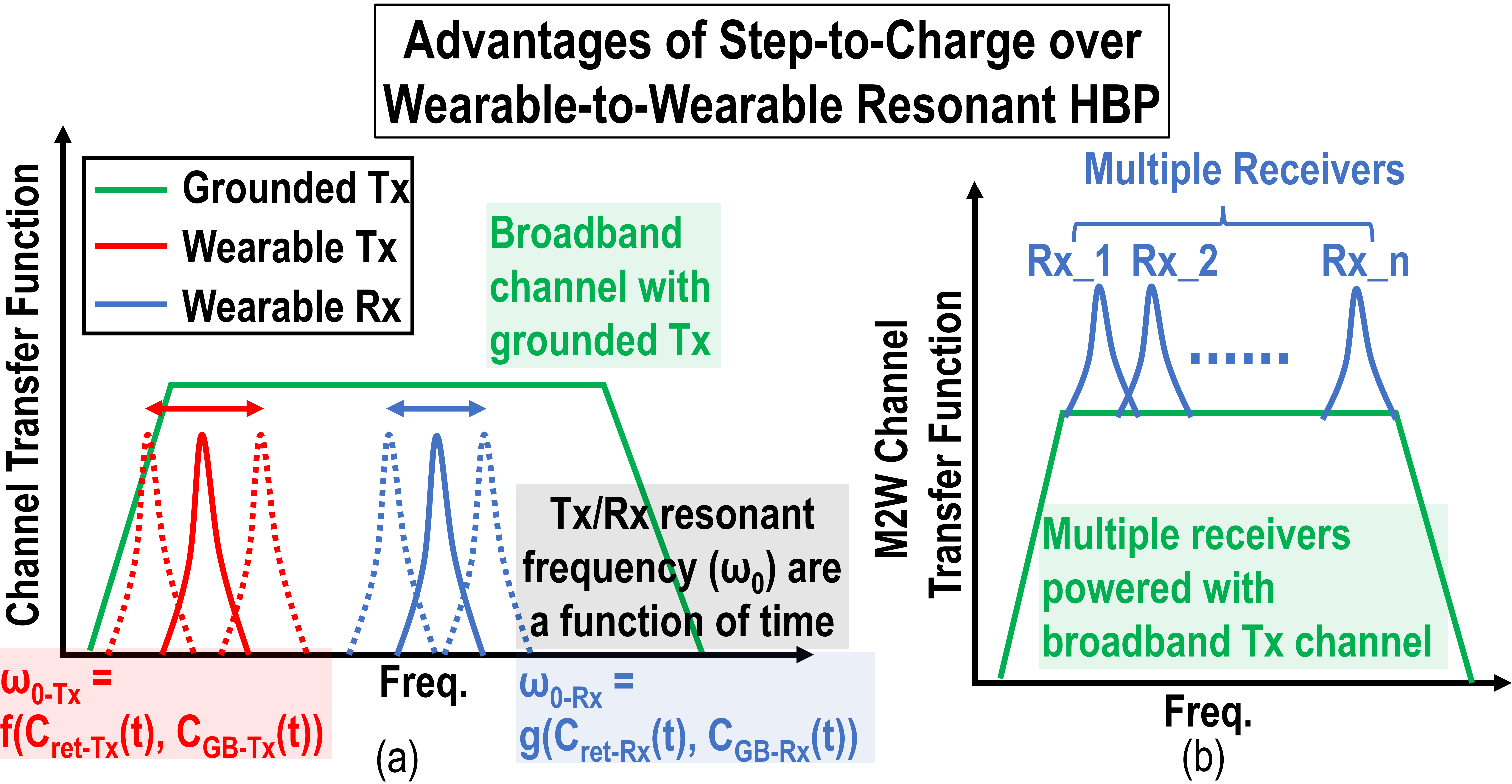}
\caption{\textit{(a) Step-to-Charge uses an earth's ground connected transmitter which provides a broadband channel across the Electro-Quasistatic frequency regime. Step-to-Charge uses only a receiver-side resonance. Wearable-to-Wearable Resonant Human Body Powering methodology requires us to tune the resonant frequencies of the transmitter and receiver to match up for optimal power delivery, which gets very difficult as the resonant frequencies are a function of time. (b) Step-to-Charge's broadband channel allows the charging of multiple receivers, which is not possible with a conventional Wearable-to-Wearable Resonant Human Body Powering system.}}
\label{fig:S2C_advantage}
\end{figure*}
%________________________________________________________________________
Step-to-Charge uses a Machine-to-Wearable Resonant Human Body Powering modality. We demonstrate the advantages of Machine-to-Wearable (M2W) modality over conventional Wearable-to-Wearable (W2W) Resonant Human Body Powering (HBP) powering.\par
A wearable transmitter (powering source) provides a high channel loss between the transmitted voltage and the on-body voltage, as shown by Datta et al. \cite{datta2021advanced}. For a wearable transmitter without the use of resonance, the on-body voltage ($V_B$) can be mathematically represented as a function of the transmitted voltage ($V_{Tx}$), the return path capacitance ($C_{ret-Tx}$), and the body capacitance ($C_{B}$) as shown in Eqn. \ref{eqn:V_B_W2W}.
\begin{equation}
    V_B = V_{Tx}\frac{C_{ret-Tx}}{C_B}
    \label{eqn:V_B_W2W}
\end{equation}
Typically, $C_{ret-Tx}$ is less than $1pF$ for wearable devices, and $C_B$ is around $100-150 pF$, making the on-body voltage more than $100$ times less than the transmitted voltage \cite{maity2018bio}.
\begin{equation}
    V_B \leq V_{Tx} \times \frac{1}{100}
\end{equation}
This additional channel loss from the transmitter to the on-body voltage allows the coupling of very low power onto the body when a wearable transmitter is used. Thus, an inductor is introduced on the transmitter side to couple a higher voltage onto the body at a particular frequency $\omega_{0-Tx}$. Using resonance on the transmitter side, the on-body voltage gets boosted by a factor ``Q" (Eqn. \ref{eqn:V_B_Res_W2W}) \cite{9693115}. 
\begin{equation}
    V_B = V_{Tx}\times Q \times \frac{C_{ret-Tx}}{C_B}
    \label{eqn:V_B_Res_W2W}    
\end{equation}
However, resonance on the transmitter side makes the wearable transmitter a narrowband source, as shown by the red curve in  Fig. \ref{fig:S2C_advantage} (a). We further employ resonance on the receiver end to increase the channel gain from the body to the receiver (blue curve in Fig. \ref{fig:S2C_advantage} (a)). The resonant frequencies of the transmitter and the receiver ($\omega_{0-Tx}$ and $\omega_{0-Rx}$) are a function of different parasitic capacitances of transmitters ($C_{ret-Tx}, C_{GB-Tx}$) and receivers ($C_{ret-Rx}, C_{GB-Rx}$). Further, as each of these parasitic capacitances is a function of the placement of the device on the body, posture of the person, and the environmental surroundings, all of which are subject to change over time, thus, the resonant frequencies of the transmitter and receiver keep changing over time. For optimal power delivery in a Wearable-to-Wearable Resonant Human Body Powering, we must ensure that the resonant frequency of the transmitter and receiver are the same (Eqn. \ref{eqn:Res_Freq}). Aligning both the resonant frequencies (transmitter and the receiver) dynamically is almost impossible, as the slightest movement of the person wearing the devices changes the parasitic capacitances.
\begin{equation*}
    \omega_{0-Tx} = \omega_{0-Rx}
\end{equation*}
\begin{equation}
    f(C_{ret-Tx}(t),C_{GB-Tx}(t)) = g(C_{ret-Rx}(t),C_{GB-Rx}(t)) 
    \label{eqn:Res_Freq}
\end{equation}

Using an earth's ground connected transmitter allows the use of a broadband powering source. As illustrated by the green curve in Fig. \ref{fig:S2C_advantage} (a). The broadband channel provided by Step-to-Charge is possible due to the high on-body voltage (Supplementary Note 2 and Supplementary Figure 2) with an earth's ground connected transmitter (Eqn. \ref{eqn:V_B_M2W}).
\begin{equation}
    V_B \approx V_{Tx}
    \label{eqn:V_B_M2W}
\end{equation}
Thus, using resonance at the receiver end doesn't require us to tediously tune the resonant frequency of the receiver due to the broadband high-channel gain possible in a Step-to-Charge M2W Resonant Human Body  Powering system. Changing the resonant frequency of the receiver ($\omega_{0-Rx}$) due to changes in posture, changing position of devices, or a change in the surroundings does not affect the power transferred to the receiver when using a Step-to-Charge system. \par
Using a broadband powering source in Step-to-Charge is at the cost of reducing powering transfer efficiency, as a higher power is transmitted to power a single receiver than a conventional W2W Resonant HBP system. However, a person typically wears multiple wearables that need to be charged. For a W2W Resonant HBP system, it is impossible to transfer power to multiple receivers, as the transmitter and receiver need to be tuned to the same frequency for any meaningful power transfer. However, with a broadband earth's ground connected powering source, it is possible to charge multiple receivers simultaneously, as illustrated by Fig. \ref{fig:S2C_advantage} (b). This also implies that the power transfer efficiency increases with more devices being charged simultaneously.

\newpage
\section*{Supplementary Note 4: Demonstration Video}\label{sec1}

The operation of Step-to-Charge is demonstrated using off-the-shelf components to power a hand-held temperature and humidity sensor, ThermoPro ``TP50". The demonstration video is available publicly on GitHub \cite{S2C_GitHub}. \par
The Step-to-Charge setup is illustrated using an Aluminum foil on the floor acting as the signal plate for the transmitter. A man standing on the Aluminum foil acts as the body channel for power transfer. The ThermoPro ``TP-50" \cite{ThermoPro}, a commercially available temperature and humidity sensing device, is housed in a box which acts as the receiver to demonstrate power transfer. The receiver setup consists of a full wave bridge rectifier using Schottky diodes (1N5819) to rectify the floor-based transmitter's AC power sent through the body. The temperature and humidity sensor is connected directly to the rectifier. Hence, the setup uses instantaneous power available from the powering source to power on the receiver.\par
%________________________________________________________________________
\begin{figure*}[b]
\centering
\includegraphics[width=\textwidth]{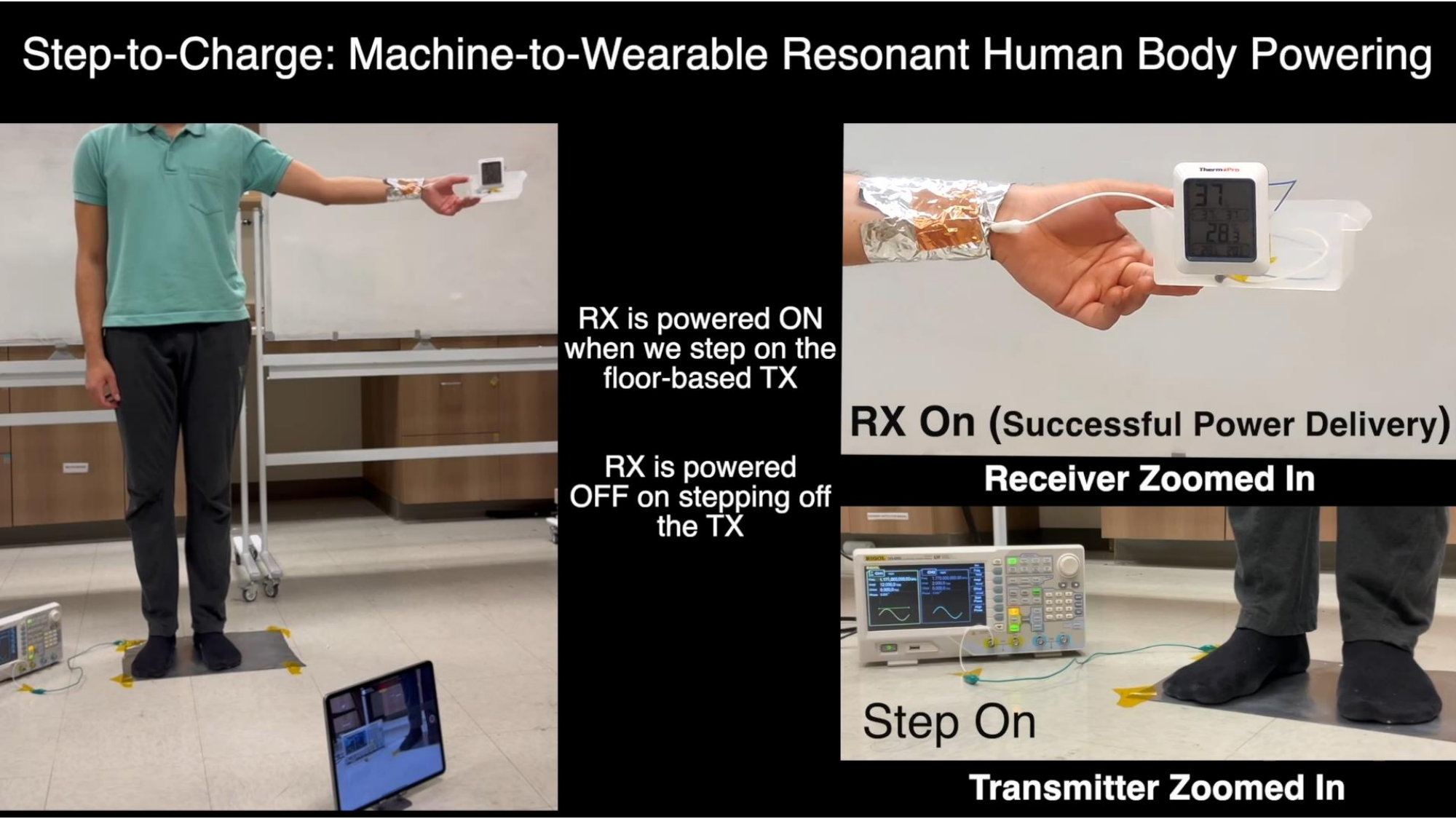}
\caption{\textit{A demonstration of Step-to-Charge, showing successful power delivery for a hand-held temperature and humidity sensor (ThermoPro TP50).}}
\label{fig:S2C_demo}
\end{figure*}
%________________________________________________________________________
It can be observed that stepping on the Aluminum foil sheet powers the wearable sensor on, thus displaying the ambient temperature and humidity. The ThermoPro TP-50 device flickers on and off initially while the person is standing on the Step-to-Charge platform due to the use of instantaneous power in powering on the device. As the slightest movement of the arm changes the parasitic capacitances, the resonant frequency keeps shifting. Thus, the power delivered to the device changes with slight changes in the position of the arm when the person moves on and off the platform. Further, the demonstration setup does not use any energy storage element. Typical energy harvesting systems use an energy storage element, like a supercapacitor, which allows the supply of constant power regardless of the available power from the transmitter. Thus, if a super capacitor is used, the power stored can be used to power on the device for an extended period of time without the device turning on intermittently due to fluctuating levels of instantaneous available power. However, in this demonstration, we illustrate that the instantaneous power available to the receiver is high enough to power on the device as soon as the person steps on the Step-to-Charge platform.\par
Fig. \ref{fig:S2C_demo} shows the demonstration setup used. The image on the left shows the complete setup, with the person moving on and off the Step-to-Charge platform to power the device on or off. The image on the top right is a zoomed in version of the receiver, and the figure on the bottom right is a zoomed in version of the transmitter. The transmitter is a wall connected Rigol ``DG4202" function generator \cite{Function_generator}, with the signal plane being an Aluminum foil sheet on the floor. This demonstration illustrates successful power transfer for a channel length of $>1m$, allowing whole body powering with EQS Resonant Human Body Powering using Step-to-Charge.

\newpage

\backmatter
\section*{Data Availability}
The data supporting the plots within this paper and other study findings are available from the corresponding author upon reasonable request.
\section*{Code Availability}
Custom codes used to process the data are available from the corresponding
author upon reasonable request.
\section*{Acknowledgments}
This work was supported by Quasistatics, Inc. – Grant 40003567, Account
F.00127126.02.036. The authors thank Dr. Shovan Maity, graduated Ph.D. student from Sparclab, Sarthak Antal, Samyadip Sarkar and David Yang, PhD students at Sparclab for their valuable input during the experimental process. 
\section*{Author contributions statement}
S.S., A.D., and L.D. conceived the idea. A.D. and L.D. conducted the theoretical analysis, FEM-based EM Simulations, and performed the experiment(s). All authors analyzed the results and reviewed the manuscript.
\section*{Competing Interests}
The authors declare that S.S. has a financial interest in Quasistatics, Inc.

\begin{comment}
\section*{Declarations}

\begin{appendices}

\section{Section title of first appendix}\label{secA1}    
\end{appendices}
\end{comment}

\bibliography{references}% common bib file

\end{document}